%% file: AoP.tex
\documentclass[conference]{IEEEtran}
\usepackage{cite}
\usepackage{amsmath}
\usepackage{algorithmic}
\usepackage{array}
\usepackage{booktabs}
\usepackage{array}
\usepackage{tabularx}
\usepackage{multirow}
\usepackage{graphicx}
\usepackage{amssymb}
\usepackage{amsthm}
\usepackage{verbatim}
\usepackage{algorithm}
\usepackage{algorithmic}
\usepackage{bm}
\usepackage{booktabs}
\usepackage{subfigure}
\usepackage{url}
\newtheorem{mythm}{Theorem}

\begin{document}
	
	\title{Age of Processing: Age-driven Status Sampling and Processing Offloading 	for Edge Computing-enabled Real-time IoT Applications}
	\author{\IEEEauthorblockN{Rui~Li, Qian~Ma, Jie~Gong, Zhi~Zhou, and Xu~Chen}\\
Sun Yat-sen University, Guangzhou, China\\
		}
	\maketitle
	
	\begin{abstract}
		The freshness of status information is of great importance for time-critical Internet of Things (IoT) applications. A metric measuring status freshness is the \textit{age-of-information} (AoI), which captures the time elapsed %of the newest received status update since it generated at the source node (e.g., a sensor)
		from the status being generated at the source node (e.g., a sensor) to the latest status update. However, in intelligent IoT applications such as video surveillance, the status information is revealed after some computation-intensive and time-consuming data processing operations, which would affect the status freshness. In this paper, we propose a novel metric, \emph{age-of-processing} (AoP), to quantify such status freshness, which captures the time elapsed of the newest received processed status data since it is generated. Compared with AoI, AoP further takes the data processing time into account. Since an IoT device has limited computation and energy resource, the device can choose to offload the data processing to the nearby edge server under constrained status sampling frequency. We aim to minimize the \emph{average} AoP in a long-term process by jointly optimizing the status sampling frequency and processing offloading policy. We formulate this online problem as an infinite-horizon constrained Markov decision process (CMDP) with average reward criterion. We then transform the CMDP problem into an unconstrained Markov decision process (MDP) by leveraging a Lagrangian method, and propose a Lagrangian transformation framework for the original CMDP problem. Furthermore, we
		\begin{comment}
		assume that the underlying Markov chain of the CMDP problem has an \textit{unichain} structure, whose optimal policy is shown to be a randomized mixture of two deterministic policies. We then develop an algorithm to obtain the optimal policy for the CMDP problem by throwing a biased coin between two deterministic policies. Simulation results illustrate the structural properties of the optimal policy, and 	
		\end{comment}
		integrate the framework with perturbation based refinement for achieving the optimal policy of the CMDP problem. Extensive numerical evaluations show that the proposed algorithm outperforms the benchmarks, with an average AoP reduction up to 30\%.\\
	\end{abstract}
	
	
	\IEEEpeerreviewmaketitle
	
	\input{Introdution}

	\input{Related_Work}
	
	\input{Problem_Formulation}
	
	\input{Algorithm1}

	\input{Algorithm2}

	\input{Simulation}

	\input{Conclusion}

	\bibliographystyle{IEEEtran}
	\bibliography{bibliography}
	
\end{document}

%% file: Introdution.tex
\section{Introduction}	

The rapid proliferation of the Internet of Things (IoT) devices boosts the fast development of various networked monitoring and cyber-physical systems applications  \cite{atzori2010internet}, \cite{8793011}, such as crowdsourcing in sensor networks \cite{DBLP:journals/corr/abs-1902-06149}, phaser updating in smart grid systems \cite{5549870}, and autonomous driving in smart transportation systems \cite{8555643}. For these IoT applications, the freshness of status information of the physical process at the operation nodes is of fundamental importance for accurate monitoring and controlling.

\emph{Age of information} (AoI), which is also often referred to as \emph{age}, was proposed to quantify the status freshness of interested physical process \cite{kaul2011minimizing}, \cite{NET-060}. More specifically, AoI is generally defined as the time elapsed from the generation at the source node (e.g., a sensor) to the last successfully received status update at the destination (e.g., a controller). \begin{comment}
	Formally, at any time $t$, if the freshest status update received at the destination was generated at time $u(t)$, then the age at the destination node is 
	\begin{equation}
	\Delta(t) = t - u(t).
	\end{equation}
\end{comment}
There have been extensive works that focus on minimizing the age under various queueing models \cite{moltafet2019age, talak2019age, akar2019finding, DBLP:journals/corr/abs-1906-12278, DBLP:journals/corr/abs-1901-08197, DBLP:journals/corr/abs-1905-13743, DBLP:journals/corr/abs-1901-10463, 8445909, DBLP:journals/corr/abs-1806-09396, DBLP:journals/corr/abs-1804-06139, 8406909, DBLP:journals/corr/abs-1801-04068, DBLP:journals/corr/abs-1709-04956, 8006592, DBLP:journals/corr/abs-1805-12586}. It is worth noting that the AoI minimization depends on the status update frequency, and differs from the conventional design principles (e.g., providing low delay). Specifically, on the one hand, updating status at a low frequency results in a small message queueing delay since the queue is always empty, however, the destination node has a large age because of the infrequent status update. On the other hand, updating status at a high frequency results in a large queueing delay due to the Little's law \cite{10.1287/opre.9.3.383}, and the destination node also has a large age because the status update suffers from a large queueing delay. Therefore, different from the queueing delay that increases with the status sampling frequency, AoI exhibits more complex patterns as a metric for status freshness and is more challenging to optimize \cite{6195689}.

For many intelligent real-time IoT applications, the status freshness depends not only on the status update frequency of AoI, but also on status data processing operations. For example, in smart video surveillance, the status update (e.g., sampling an image) would not take effect until the useful information embedded in the image is extracted by some data processing operations (e.g., AI-based image recognition) which are computational expensive and time consuming. 

Since an IoT device typically has limited computation and storage capacities, edge computing can be leveraged to facilitate real-time data processing. In this case, the IoT device can offload the data processing operations to the nearby mobile edge computing (MEC) platforms \cite{8815809}, which utilize the edge servers deployed at the edge of radio access networks (e.g., base stations (BSs) or access points (APs)) to execute computing tasks. Specifically, the IoT device offloads the status update to the edge server through wireless channel for further data processing, and then the edge server sends the final results back to the destination node. Therefore, the processing offloading would also affect the status freshness. 

To capture the status freshness considering data processing in the edge computing-enabled real-time IoT applications, we propose a new metric, \textit{age-of-processing} (AoP), which is defined as the time elapsed since the generation of the freshest status update data until it is processed and  finally takes effect at the destination node. Compared with conventional AoI, the AoP takes the additional data processing time in status update into account. 

In this paper we aim to minimize the AoP through optimizing the data processing offloading decision and the status sampling frequency jointly. Specifically, the data processing offloading can reduce the data processing time by utilizing edge servers' computation resource, but incurs additional transmission time which depends on the wireless channel state between the source node (e.g., the IoT device) and the edge server. When the wireless channel state is good, offloading the data processing operations to the edge server incurs short transmission time and can reduce the processing time. However, when the channel state is bad, the transmission time between the source node and the edge server is not negligible, and the IoT device can process the status update data by its local server or wait for a good channel state. Therefore, we need to carefully decide the optimal offloading strategy under different channel states to minimize the AoP. 

Moreover, the status sampling frequency also has an essential impact on AoP. Specifically, when the previous status update is under processing, a new update needs to wait in queue, and hence becomes stale while waiting. Therefore, it can be better not to generate new sample while the edge server is busy. Authors in \cite{8684949} proposed a status sampling policy called \textit{zero-wait} policy, which samples a new update after the previous update takes effect. However, authors in \cite{7283009}, \cite{7524524} showed that the zero-wait policy might be far from age-optimal in some cases. Hence, how to optimize the status sampling frequency considering data processing is still an open question. Furthermore, the status sampling process consumes energy of IoT devices. It is necessary to introduce a constraint for the sampling frequency due to the limited energy budget of the IoT devices, which make it harder to obtain the optimal status sampling policy for minimizing the AoP. 
\begin{comment}
	Although a higher sampling frequency can achieve a lower average AoP due to the more frequent status update, it will run out the energy of the IoT device. Therefore, we need to carefully balance the trade-off between the AoP reduction and energy consumption.
\end{comment}

By addressing the challenges above, we achieve the following key contributions: 
\begin{enumerate}
	\item We propose a new metric, \textit{age-of-processing} (AoP), to capture the status freshness considering data processing in real-time IoT applications. In order to minimize the average AoP, we formulate the joint status sampling and processing offloading problem as an infinite-horizon constrained Markov decision process (CMDP) with the maximum sampling frequency constraint of the IoT device. 
	
	\item We relax the challenging CMDP problem into an unconstrained MDP problem using the Lagrangian method which significantly simplifies the original CMDP problem. We then propose a Lagrangian transformation framework to derive the optimal status sampling and processing offloading policy under the optimal Lagrangian multiplier. 
	
	\item Building upon the proposed Lagrangian transformation framework, we develop stochastic approximation based policy iteration algorithms with perturbation based refinement to achieve the optimal policy of the CMDP problem.

	\item We provide extensive simulations to illustrate the structural properties of the optimal policy, and show that the proposed improved algorithm outperforms the benchmarks, with an average AoP reduction up to 30\%.
\end{enumerate}

The rest of the paper is organized as follows. In Sec. II, we discuss the related works. In Sec. III, we present the system model and formulate the AoP minimization problem as a CMDP problem. In Sec. IV, we transform the CMDP problem to an unconstrained MDP problem by leveraging the Lagrangian method. In Sec V, we first propose a Lagrangian transformation framework for the original CMDP problem, and improve it with perturbation based refinement to achieve the optimal policy. We show our simulation results in Sec. VI, and conclude the paper in Sec.VII.

%% file: Related_Work.tex
\section{Related work}

Age-of-information (AoI) was introduced in the early 2010s as a new metric to characterize the freshness of the information that a system has about a process observed remotely \cite{kaul2011minimizing}. Since then, an abundant of researches focus on the queueing theory to analyze the age-of-information in various system settings. In \cite{6195689}, the authors obtained the theoretical results of the average AoI, where the status update is served with the first-come-first-served (FCFS) principle, and more specifically, the queueing models include $M/M/1$, $M/D/1$ and $D/M/1$. After that, different queueing models, such as $G/G/1$ \cite{DBLP:journals/corr/abs-1905-13743}, $M/G/1$ \cite{8445909}, and $D/G/1$ \cite{8406909}, were also studied. A new metric, peak age, was introduce in \cite{6875100}, and the authors in \cite{akar2019finding} obtained the distribution of peak age in a $PH/PH/1/1$ queue. In \cite{DBLP:journals/corr/abs-1806-09396}, the authors studied the reliable transmission under the peak-age violation guarantees.

Another branch of researches on AoI considers energy-harvesting constraints since the IoT device (e.g., a sensor) is usually energy limited, and the sampling process consumes energy \cite{DBLP:journals/corr/abs-1907-03826, ceran2019reinforcement, Arafa_2018, DBLP:journals/corr/abs-1802-02129}. In \cite{DBLP:journals/corr/abs-1907-03826}, the authors derived an optimal transmission policy in an energy harvesting status update system, which is formulated as an MDP problem. In \cite{ceran2019reinforcement}, the authors proposed a reinforcement learning algorithm to learn the system parameters and the status update policy for an energy harvesting transmitter with a finite-capacity battery. The authors in \cite{Arafa_2018}, \cite{DBLP:journals/corr/abs-1802-02129} analyzed the scenario where an energy harvesting sensor with random or incremental battery recharge sends measurement updates to a destination, and showed that the optimal update policy follows a renewal structure. All the above works assume that the status update takes effect once it is received in the destination node, and the age is immediately set down to the time elapsed from status generation to its reception. 

For computation-intensive application (e.g., autonomous driving), however, the status update (e.g., a video clip) needs further data processing to reveal the useful features. Hence, the data processing time also affects the age. However, there are very limited research efforts in this area. In \cite{8635855}, the authors considered the soft update in an information updating system. In both exponentially and linearly decaying age cases, the authors derived the optimal sampling schemes subject to a sampling frequency constraint. In \cite{kuang2020analysis}, the authors studied the AoI for computation-intensive messages with MEC, and derived the closed-form average
AoI for exponentially distributed computing time. In \cite{DBLP:journals/corr/abs-1811-12924}, the authors jointly optimized the information freshness (age) and the completion time in a vehicular network. Nevertheless, the computation time is not taken into consideration in the age. In \cite{song2019age}, the authors proposed a performance metric called age of task (AoT) to evaluate the temporal value of computation tasks. By jointly considering task scheduling, computation offloading and energy consumption, the authors proposed a light-weight task scheduling algorithm. However, it is an offline policy where the task arrival time is known in advanced. 

\begin{comment}
The most relevant work to this paper is \cite{7524524}, in which the authors found that the zero-wait status sampling policy is not always age-optimal. They showed that the optimal status sampling policy which minimizes the AoI has a threshold structure, and proposed algorithm to obtain it. In this paper, 
\end{comment}

Different from existing research efforts, in this paper, we expand the concept of AoI to AoP by taking the data processing time into consideration. We further consider data processing offloading to MEC server, and minimize the total average AoP by optimizing the status sampling and processing offloading policy. 

%% file: Problem_Formulation.tex
\section{Model and formulation of AoP minimization}

\subsection{System Model}
Consider a real-time IoT status monitoring and control system for computation-intensive applications. The IoT device (a.k.a. the sensor) monitors the current status of a physical process (e.g., a camera records images of traffic situation at a crossroad), which needs further data processing. As shown in Fig. 1, the IoT device can choose to process the raw data locally at its processor or offload them to a mobile edge server in proximity. The data processing operation reveals the hidden feature (e.g., the congestion level at the crossroad) in the raw data, which we refer to as knowledge that will be then transmitted to an operator for accurate control.  After receiving the knowledge, the operator sends an acknowledge (ACK) to the IoT device to sample a new status update.

We define the time elapsed from the status generation at the IoT device to the latest knowledge received by the operator as the \emph{age-of-processing} (AoP), which is maintained by the operator to capture the status freshness. Compared to the traditional AoI, the AoP takes the data processing time into account, which is affected by the data processing offloading policy. 

The IoT device follows the \textit{generate-at-will} sampling policy \cite{7283009}, under which the IoT device can start a new sample whenever it prefers, and does not generate a new status update when the previous update is under processing, to avoid unnecessary waiting time. Suppose the IoT device samples a new status update $i$ at time $S_i$, and then decides where to send the raw data (e.g., to its local processor or the edge server) for further data processing.

For the status update $i$, we denote its data processing task by a pair $(l_i,c_i)$, where $l_i$ is the input data size of status packet and $c_i$ is the total required CPU cycles to compute this task.

\subsubsection{Local processing}
We assume that the sensor is equipped  with a local processor (e.g., embedded CPU) for some necessary computations. If the sensor chooses to process the status update locally, then the operation time can be formulated as 

\begin{equation}
	t_i^l = \frac{c_i}{f_l},
\end{equation}
where $f_l$ is the CPU frequency of the local processor. After data processing, \begin{comment}
the hidden feature of the raw data can be exposed, which we refer to as the knowledge of the status update. Finally,
\end{comment} 
the local server transmits the processed status result to the operator. We assume that the data size of the result is quite small (e.g., the result of object classification is usually takes only several bits). Therefore,  the time of transmitting the result to the operator is negligible.

\subsubsection{Edge offloading}

\begin{figure}[t]
	\centering
	\includegraphics[scale=0.7]{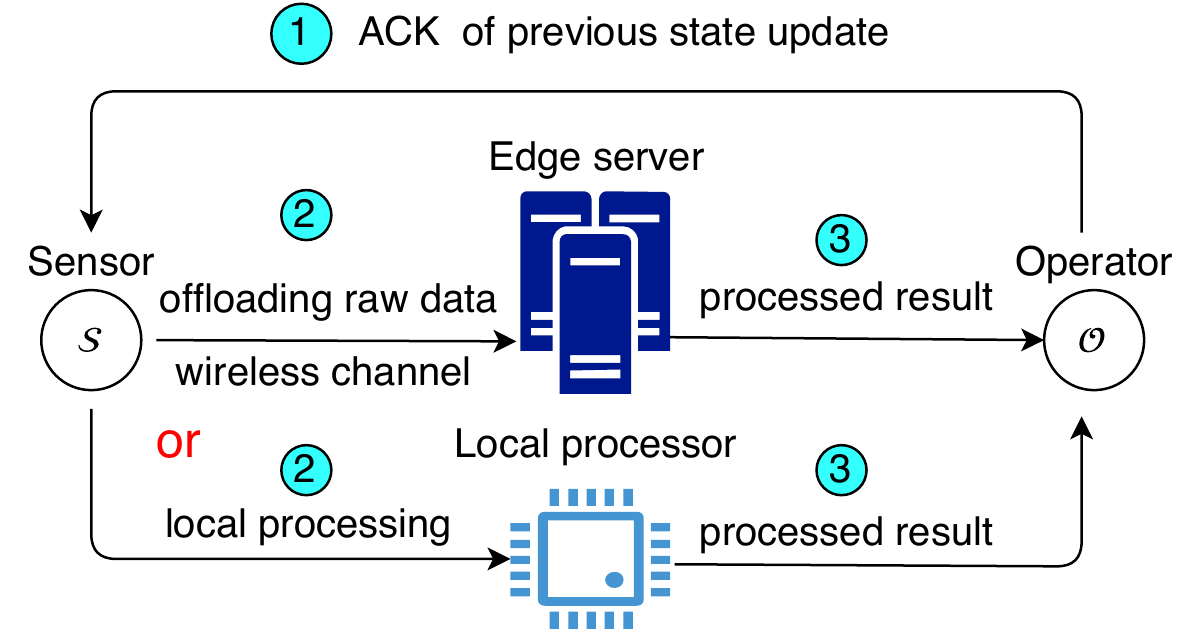}
	\centering
	\caption{Status sampling and processing procedure.}
\end{figure}

If the sensor chooses to offload the raw data to the edge server, it incurs extra time for transmitting the computation input data via wireless connection. According to \cite{rappaport1996wireless}, the offloading rate can be formulated as 

\begin{equation}
	r_i = W \log_2 \left(1+\frac{p_ih_i}{\sigma_i^2}\right), \label{3}
\end{equation} 
where $W$ is the channel bandwidth and $p_i$ is the transmission power of update $i$. Furthermore, $h_i$ denotes the wireless channel gain between the sensor and the edge server, which can be generated using a distance-dependent path-loss model given in \cite{chu2013heterogeneous}

\begin{equation}
	h_i\mathrm{[dB]} = 140.7 + 36.7\log_{10}d\mathrm{[km]},
\end{equation}
and $\sigma_i^2$ is the total background noise and interference power of transmitting update data $i$. Therefore, the transmission time is computed as 

\begin{equation}
	t_i^{tr} = \frac{l_i}{r_i}, \label{5}
\end{equation}
and the data processing time at the edge server is 

\begin{equation}
	t_i^{ex} = \frac{c_i}{f_e}, \label{6}
\end{equation}
where $f_e$ is the CPU frequency of the edge server. As mentioned before, we ignore the time of transmitting the processed statue result from the edge server to the operator. Following \eqref{5} and \eqref{6}, we can compute the time overhead of the edge offloading approach as

\begin{equation}
	t_i^e = t_i^{tr} + t_i^{ex}.
\end{equation}  

Throughout this paper, we assume that both the computation capacities of the local and edge servers are stable (e.g., $f_l$ and $f_e$ are both constants). This is reasonable since the sensor usually carries out a dedicated sensing task and the edge server usually allocates a resource block (i.e., a virtual machine) with fixed size to a certain computing task. Besides, we assume that all state update tasks have the same input data size $l_i$ and required computation $c_l$\footnote{We left the heterogeneous state update tasks with different $l_i$ and $c_i$ in our future work.}. For example, the input image size for object recognition based surveillance task is the same with almost the same CPU cycles for processing each image. For the wireless channel, we assume that the transmission power $p_i$ is the same for all update $i$. The total background noise and interference power $\sigma_i^2$ influences the wireless channel state. It is unknown and change stochastically. The channel state has a critical impact on the data offloading policy. Intuitively, when the wireless channel state is good (e.g., $\sigma_i^2$ is small), the IoT device tends to offload the status data to the edge server, using the abundant computing resource of edge server to reduce the processing time. When the wireless channel is bad (e.g., $\sigma_i^2$ is large), the transmission time is relatively large, and hence the IoT device would like to process the update data locally to avoid the large transmission time.
\begin{figure*}[t]
	\centering
	\includegraphics[scale=0.8]{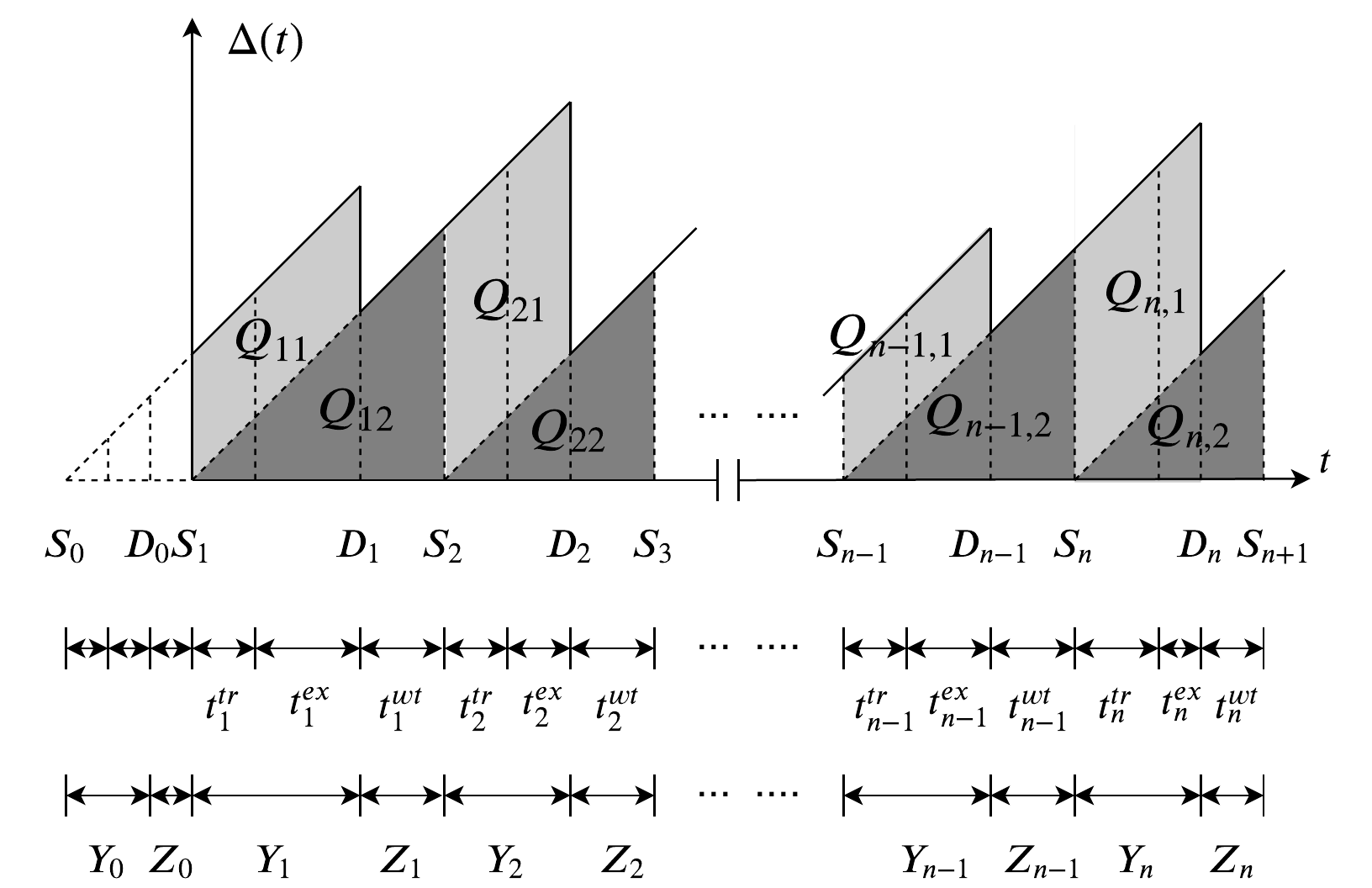}
	\centering
	\caption{Evolution of the age-of-processing (AoP).}
\end{figure*}
We depict the evolution of the age-of-processing in Fig. 2. Suppose a new status update $i$ is sampled at time $S_i$. If the raw data is processed locally, then the processing time is $Y_i = t_l$\footnote{Since all state update tasks have the same required computation $c_i$, we simplify $t^l_i$ as $t_l$ for all update $i$.}. If the raw data is processed in edge server, the total processing time is $Y_i = t_i^{tr} + t_i^{ex}$. Therefore, the processed result of update $i$ is delivered at time $D_i = S_i + Y_i$. After the operator receives the  update $i$, the sensor node may insert a waiting time $Z_i \in [0, T]$ before sampling the new status update $i+1$ at time $S_{i+1} = D_i + Z_i$, where $T$ is the maximum waiting time under a sampling frequency constraint. The sensor can switch to a low-power sleep mode during the waiting period $[D_i, S_{i+1})$. 

At any time $t$, the freshest status update at the operator is generated at time 

\begin{equation}
	u(t) = \max\{S_i:D_i \le t\}.
\end{equation}
Then the age-of-processing $\Delta(t)$ is defined as

\begin{equation}
	\Delta(t) = t - u(t).
\end{equation}
As shown in Fig. 2, the AoP $\Delta(t)$ follows a stochastic process which increases linearly with $t$ while waiting for the next sample or the data is under processing, and then downward jumps when the status update is delivered at the operator. Therefore, the curve of the AoP has a zig-zag shape. More specifically, status update $i$ is sampled at time $S_i$ and is received by the operator at time $D_i = S_i + Y_i$. Therefore, the AoP at time $D_i$ is $\Delta(D_i) = D_i - S_i = Y_i$. After that, the AoP continues to increase linearly with time $t$ while the sensor is waiting or the update data is under processing. Finally, the AoP reaches $\Delta(D_{i+1}^-) = Y_i + Z_i + Y_{i+1}$ right before the processed result of status update $i+1$ is delivered. Then, at time $D_{i+1}$, the AoP drops to $\Delta(D_{i+1})=Y_{i+1}$.

\subsection{CMDP Formulation}
In this subsection, we focus on specifying the optimal status sampling and computation offloading policy to minimize the average AoP of the system discussed above.

In most existing works, the long-term average age is a key performance metric to measure the long-run information freshness, which is defined as

\begin{equation}
	\Delta_{av} = \lim_{t \rightarrow \infty} \frac{1}{t} \int_{0}^{t} \Delta(t) \mathrm{dt}.
\end{equation}
Intuitively, $\Delta_{av}$ is the time-average shaded area under the $\Delta(t)$ envelop. To compute the average AoP, we decompose $\Delta(t)$ into a series of areas between the sampling time $S_i$. As shown in Fig. 2, the light shaded area $Q_{i1}$ is a parallelogram, which equals to

\begin{equation}
	Q_{i1} = (Y_{i-1} + Z_{i-1})Y_i,
\end{equation}
and the dark shaded area $Q_{i2}$ is a triangle having the area

\begin{equation}
	Q_{i2} = \frac{1}{2}(Y_i+Z_i)^2.
\end{equation}
Therefore, the average AoP can be calculated as 

\begin{equation}
	\begin{aligned}
	\overline{Q} &=  \frac{\sum_{i \rightarrow \infty}(Q_{i1}+Q_{i2})}{\sum_{i \rightarrow \infty}(Y_i + Z_i)}\\
	&= \frac{\sum_{i \rightarrow \infty}\big[(Y_{i-1} + Z_{i-1})Y_i+\frac{1}{2}(Y_i+Z_i)^2\big]}{\sum_{i \rightarrow \infty}(Y_i + Z_i)}
	.\end{aligned}
\end{equation}
Note that, minimizing $\overline{Q}$ is a long-term stochastic problem. At each delivered time $D_i$, the operator maintains the age $\Delta(D_i)=Y_i$, and then decides the inserted waiting time $Z_i$ before sampling next status update $i+1$. Besides, we assume that the IoT device also determines where to process the next update $i+1$ at time $D_i$, which will affect the value of $Y_{i+1}$.

\textbf{Markov decision process:} As mentioned before, the wireless channel states between the sensor and the edge server change stochastically. Let $X_i$ be the channel state with a finite state space $\Gamma = \{\gamma_0,\ldots,\gamma_M\}$,\footnote{Since the data size of all status update packets is the same, we can simplify the channel state $\gamma_0,\ldots,\gamma_M$ as the transmission time of the update data \eqref{5}. Since the transmission time in  \eqref{5} is continuous, it results in an infinite state space in the MDP. For simplicity, we discretize the transmission time into $M+1$ channel states.} which is influenced by $\sigma_i^2$. Unlike the assumption of the i.i.d. channel state process $\{X_i, i=1,2,\ldots\}$ in \cite{DBLP:journals/corr/abs-1807-04356}, we consider a general case where the process of $X_i$ is a stationary and ergodic Markov chain with the transition matrix $\mathbf{P}^{ch}$ \cite{zhang1999finite}.\footnote{We assume that we know the statistics of the channel $\mathbf{P}^{ch}$ in advance, since we can estimate $\mathbf{P}^{ch}$ through channel training.} The element $P_{ij}$ the transition matrix $\mathbf{P}^{ch}$ is the probability from channel state $\gamma_i$ to state $\gamma_j$.

At time $D_i$, we denote $A_i \triangleq \{Y_{i-1},Z_{i-1},Y_i,X_i\} \in \mathcal{A}$ as the current system state, where $\mathcal{A}$ is the system state space.\footnote{We also discretize the waiting time $Z_i$ since a discrete waiting time is much easier to execute for IoT devices, and an infinite waiting time space results in an infinite MDP state space which is difficult to solve.} Then the sensor chooses an action $\bm{w}_i \triangleq \{Z_i, O_i\} \in \mathcal{W}$ from the action space $\mathcal{W}$, where $Z_i$ is the inserted waiting time and $O_i$ is the offloading decision for update $i$. When $O_i=0$, the sensor chooses to offload the status update $i$ to the edge server, and when $O_i=1$, the sensor chooses to process the update $i$ locally. We then define the reward function for taking action $\bm{w}_i$ at state $A_i$ as

\begin{equation}
\begin{aligned}
r(A_i, \bm{w}_i)&= Q_i = Q_{i1}+Q_{i2}\\&=(Y_{i-1} + Z_{i-1})Y_i+\frac{1}{2}(Y_i+Z_i)^2.
\end{aligned}
\end{equation}
The system then evolves to the next state $A_{i+1} = \{Y_i,Z_i,Y_{i+1},X_{i+1}\}$, which only depends on previous system state $A_i$ and the action $\bm{w}_i$.  More specifically, the transition of channel state is

\begin{equation}
\mathbb{P}(X_{i+1}=\gamma_m|X_i=\gamma_j) = P_{jm},
\end{equation}
where $P_{jm}$ is the element of channel transition matrix $\mathbf{P}^{ch}$, and the age evolves according to

\begin{equation}
Y_{i+1}=\begin{cases}
t^{ex} + \gamma_m & \text{if } O_i=0, X_{i+1}=\gamma_m,\\
t_l               & \text{if } O_i=1.
\end{cases}
\end{equation}

\textbf{Stationary status sampling and processing offloading policy:} Given the system state $A \in \mathcal{A}$, the IoT device determines the sampling and offloading action $\bm{w} \in \mathcal{W}$ according to the following policy.

\textbf{Definition 1:} A \textit{stationary status sampling and computation offloading policy} $\pi$ is defined as a mapping from the system state space $\mathcal{A}$ to the control action space $\mathcal{W}$, where $\pi:\mathcal{A} \rightarrow \mathcal{W}$, which is independent of the update sequence $i$. 

In this paper, we focus on the stationary policy due to its low complexity for practical implementation (e.g., without recording the long historical information for decision making). Under a given stationary policy $\pi$, the average AoP can be calculated as:

\begin{equation}
	\overline{Q}(\pi) \triangleq \limsup_{n \rightarrow \infty}  \frac{\mathbb{E}_{\pi}\left[\sum_{i=1}^{n}Q_i\right]}{\mathbb{E}_{\pi}\left[\sum_{i=1}^{n}(Y_i+Z_i)\right]}, \label{15}
\end{equation}
where the expectation operation is taken with respect to the measure induced by the policy $\pi$, and we focus on the worst case derived by the $\limsup$ operation. 

\textbf{Sampling frequency constraint:} Due to the limited energy resource of the sensor, it is impossible to sample the status update in a very high frequency. Following the works \cite{7524524} and \cite{DBLP:journals/corr/abs-1710-04971}, we introduce a sampling frequency constraint 

\begin{equation}
	\overline{T}(\pi) \triangleq \liminf_{n \rightarrow \infty}  \frac{1}{n}\mathbb{E}_{\pi}\left[\sum_{i=1}^{n}(Y_i+Z_i)\right] \ge T_{min}, \label{16}
\end{equation}
where $T_{min}=1/f_{max}$ is the minimum sampling duration and $f_{max}$ the maximum allowed average status sampling frequency due to a long-term average resource constraint. We should emphasize that in practice it is hard to monitor the runtime energy expenditure by the sensor itself, and hence we consider the maximum sampling frequency constraint instead of the energy budget constraint in the formulation. 

\textbf{AoP minimization:} We seek to find the optimal stationary status sampling and computation offloading policy $\overline{\pi}^{*}$ that minimizes the average AoP under a maximum sampling frequency constraint at the sensor, as follows:

\begin{equation}
\begin{aligned}
&\overline{Q}^{*} \triangleq \min_{\pi}\overline{Q}(\pi),\\
&\mathrm{\textbf{s. t. }} \overline{T}(\pi) \ge T_{min}. \label{17}
\end{aligned}
\end{equation}
Problem \eqref{17} is a constrained Markov decision process (CMDP). It is computationally intractable to find the optimal policy $\overline{\pi}^{*}$ for problem \eqref{17}, since only at the end of the infinite trajectory can we obtain the final valuation $\overline{Q}(\pi)$ of the policy $\pi$, this is because the denominator of \eqref{15} is the sum of $Y_i+Z_i$ for all status update $i$. 

To tackle this difficulty, we relax the problem \eqref{17} as:

\begin{equation}
\begin{aligned}
&\widetilde{Q}^{*} \triangleq \min_{\pi}\widetilde{Q}(\pi),\\
&\mathrm{\textbf{s. t. }} \overline{T}(\pi) \ge T_{min}, \label{18}
\end{aligned}
\end{equation}
where

\begin{equation}
\widetilde{Q}(\pi) \triangleq \limsup_{n \rightarrow \infty} \frac{1}{n}\mathbb{E}_{\pi}\left[\sum_{i=1}^{n}\widetilde{Q}_i\right], \label{19}
\end{equation}
and 

\begin{equation}
	\widetilde{Q}_i = \frac{Q_i}{Y_i+Z_i} = \frac{Y_{i-1} + Z_{i-1}}{Y_i+Z_i}Y_i + \frac{1}{2}(Y_i+Z_i).
\end{equation}
Obviously, finding the optimal policy $\widetilde{\pi}^{*}$ for problem \eqref{18} is not equal to the optimal policy $\overline{\pi}^{*}$ for problem \eqref{17}. If $\overline{Q}(\pi)$ is smaller than $\widetilde{Q}(\pi)$ for all policy $\pi$, therefore, the solution of problem \eqref{18} is an upper bound policy for the original problem \eqref{17}. However, there is no certain assertion to determine the direction of inequality between 

\begin{equation*}
	 \frac{\sum_{i=1}^{n}Q_i}{\sum_{i=1}^{n}(Y_i+Z_i)} \quad \mathrm{and} \quad \frac{1}{n}\sum_{i=1}^{n}\frac{Q_i}{Y_i+Z_i}.	 
\end{equation*}
The inequality direction is influenced by the values of $n$ and all $Q_i/(Y_i+Z_i)$. However, the extensive simulation results in Sec. VI show that the ratio between $\overline{Q}(\widetilde{\pi}^{*})$ and $\widetilde{Q}(\widetilde{\pi}^{*})$ is very close to 1, which shows that the relaxed problem \eqref{18} is a good approximation to the original problem \eqref{17}.

%% file: Algorithm1.tex
\section{Unconstrained MDP Transformation}
It is well known that solving a CMDP problem directly is quite challenging \cite{altman1999constrained}. In this section we will transform the CMDP problem \eqref{18} to an unconstrained MDP problem by leveraging the Lagrangian method. 

We first describe problem \eqref{18} in terms of CMDP. At each delivered time $D_i$ which we also refer to as decision epoch $i$, the IoT device observes the current system state $A_i = \{Y_{i-1},Z_{i-1},Y_i,X_i\}$, where $Y_{i-1}$ is the processing time of the previous status update $i-1$, $Z_{i-1}$ is the waiting time before sampling the update $i$, and $Y_i$, $X_i$ are current processing time and transmission time, respectively. After observing the current state $A_i$, the IoT device selects an action $\bm{w}_i$ following a policy $\pi$, where $\bm{w}_i=\pi(A_i)$. We also refer the policy $\pi$ to a state-action mapping function. After that, the IoT device will receive an immediate reward $\widetilde{Q}_i$ from the reward function

\begin{equation}
	\widetilde{r}(A_i, \bm{w}_i) = \widetilde{Q}_i = \frac{Y_{i-1} + Z_{i-1}}{Y_i+Z_i}Y_i + \frac{1}{2}(Y_i+Z_i),
\end{equation}
which is a time-average area of $Q_i$ and then the system evolves to next state $A_{i+1} = \{Y_i,Z_i,Y_{i+1},X_{i+1}\}$. We can see that all the elements in $A_{i+1}$ only depend on the previous state $A_i$ and action $\bm{w}_i$. Therefore, the random process $\{A_i\}$ is a \textit{controlled Markov process}. The objective of problem \eqref{18} is to find an optimal state-action mapping function $\widetilde{\pi}^{*}$ to minimize the infinite horizon average reward

\begin{equation}
	\widetilde{Q}(\widetilde{\pi}^{*}) = \min_{\pi}\limsup_{n \rightarrow \infty} \frac{1}{n}\mathbb{E}_{\pi}\left[\sum_{i=1}^{n}\widetilde{Q}_i\right],
\end{equation}
while committing to a sampling constraint $\overline{T}(\pi) \ge T_{min}$.

%We recall that problem \eqref{18} is a constrained Markov decision process (CMDP) with average reward criterion.
A major challenge in obtaining the optimal policy for problem \eqref{18} is the sampling  frequency constraint. To overcome this difficulty, we first transform the problem \eqref{18} into an unconstrained Markov decision process (MDP) by introducing \textit{Lagrange multipliers} \cite{DBLP:journals/corr/abs-1807-04356}.  
We define the immediate Lagrange reward of update $i$ as

\begin{equation}
	L^{\lambda}(A_i, \bm{w}_i) \triangleq \widetilde{r}(A_i, \bm{w}_i) - \lambda(Y_i+Z_i),
\end{equation}
where $\lambda \ge 0$ is the Lagrange multiplier. Then, the average Lagrange reward under policy $\pi$ is given by

\begin{equation}
\overline{L}^{\lambda}(\pi) \triangleq \limsup_{n \rightarrow \infty}\frac{1}{n}\mathbb{E}_{\pi}\left[\sum_{i=1}^{n}L^{\lambda}(A_i, \bm{w}_i)\right].
\end{equation}
By introducing the Lagrange multiplier, we now have an unconstrained MDP problem with the objective of minimizing the average Lagrange cost

\begin{equation}
	\overline{L}^{\lambda}(\pi^{\lambda}) \triangleq \min_{\pi}\overline{L}^{\lambda}(\pi). \label{27}
\end{equation}
Let $\pi^{\lambda}$ be the optimal policy of problem \eqref{27} when the Lagrange multiplier is $\lambda$. Define
$\overline{L}^{\lambda}=\overline{L}^{\lambda}(\pi^{\lambda})$, $\widetilde{Q}^{\lambda}=\widetilde{Q}^{\lambda}(\pi^{\lambda})$, and $\overline{T}^{\lambda}=\overline{T}^{\lambda}(\pi^{\lambda})$. For the above Lagrange transformation, we can show the following result. 

\textbf{\textit{Lemma}} 1: $\overline{L}^{\lambda}$  is monotone non-increasing while $\widetilde{Q}^{\lambda}$ and $\overline{T}^{\lambda}$ are monotone non-decreasing in $\lambda$.
\begin{proof}
	The monotone non-increasing property of  $\overline{L}^{\lambda}$ and non-decreasing property $\widetilde{Q}^{\lambda}$ are a consequence of the following fundamental inequality
	\begin{equation}
		\begin{aligned}
		\overline{L}^{\lambda+\eta}(\pi^{\lambda+\eta})-\overline{L}^{\lambda}(\pi^{\lambda+\eta}) &\le \overline{L}^{\lambda+\eta}(\pi^{\lambda+\eta}) - \overline{L}^{\lambda}(\pi^{\lambda})\\
		&\le \overline{L}^{\lambda+\eta}(\pi^{\lambda}) - \overline{L}^{\lambda}(\pi^{\lambda}) \le 0,
		\end{aligned}
	\end{equation}
	for any positives $\lambda \ge 0$ and $\eta > 0$. The first inequality follows that the policy $\pi^{\lambda}$ minimizes the problem $\overline{L}^{\lambda}(\pi)$, and the second inequality follows that the policy $\pi^{\lambda+\eta}$ minimizes the problem $\overline{L}^{\lambda+\eta}(\pi)$. The third inequality can be obtained  from
	\begin{equation}
		-\eta\overline{T}^{\lambda+\eta} \le \overline{L}^{\lambda+\eta}-\overline{L}^{\lambda} \le -\eta\overline{T}^{\lambda} \le 0.
	\end{equation}
	Therefore, we have $\overline{L}^{\lambda} \ge \overline{L}^{\lambda+\eta}$ and $\overline{T}^{\lambda} \le \overline{T}^{\lambda+\eta}$. As for $\widetilde{Q}^{\lambda}$, we first assume that $\widetilde{Q}^{\lambda}$ is not monotone non-decreasing. Then there exists $\lambda$, $\eta$ such that $\widetilde{Q}^{\lambda} > \widetilde{Q}^{\lambda+\eta}$. But $\overline{T}^{\lambda} \le \overline{T}^{\lambda+\eta}$, whence,
	\begin{equation}
		\widetilde{Q}^{\lambda}-\lambda\overline{T}^{\lambda} > \overline{Q}^{\lambda+\eta} - \lambda\overline{T}^{\lambda+\eta}.
	\end{equation}
	Consequently, we come to the contradiction $\overline{L}^{\lambda}(\pi^{\lambda}) > \overline{L}^{\lambda}(\pi^{\lambda+\eta})$. Finally, we have the result $\widetilde{Q}^{\lambda} \le \widetilde{Q}^{\lambda+\eta}$.
\end{proof}
Lemma 1 reveals important relationships between the Lagrange multiplier $\lambda$ and the minimum sampling duration $\overline{T}(\pi)$ as well as the average AoP $\widetilde{Q}(\pi)$, which help us solve the MDP problem \eqref{27}. First, the minimum sampling duration  $\overline{T}(\pi)$ is non-decreasing in $\lambda$. Therefore, the optimal policy $\pi^{\lambda}$ to problem \eqref{27} under Lagrange multiplier $\lambda$ corresponds to a certain $\overline{T}(\pi^{\lambda})$. When $\overline{T}(\pi^{\lambda}) \le T_{min}$, the policy $\pi^{\lambda}$ is not a feasible solution to the original problem \eqref{17}. Then, we can increase the value of $\lambda$, until $\overline{T}(\pi^{\lambda}) \ge T_{min}$. Furthermore, the average AoP, $\widetilde{Q}(\pi^{\lambda})$ is also non-decreasing in $\lambda$. Since our objective is to find an optimal policy $\widetilde{\pi}^{*}$ to minimize $\widetilde{Q}(\pi)$ subject to $\overline{T}(\pi) \ge T_{min}$, it is equivalent to find the optimal Lagrange multiplier $\lambda^{*}$, such that
\begin{equation}
	\lambda^{*} = \inf \{\lambda:\overline{T}(\pi^{\lambda}) \ge T_{min}\}. \label{31}
\end{equation}
In order to find the optimal Lagrange multiplier $\lambda^{*}$, we need to solve the following two subproblems:

\textbf{Subproblem 1:} how to find the optimal policy $\pi^{\lambda}$ for the MDP problem \eqref{27} when given a Lagrange multiplier $\lambda$;

\textbf{Subproblem 2:} how to update $\lambda$ such that $\lambda$ converges to $\lambda^{*}$.

In summary, the Lagrangian transformation method transforms the CMDP problem \eqref{18} to the unconstrained MDP problem \eqref{27} which is much easier to solve. Furthermore, by exploring the relationships between the Lagrangian multiplier and the sampling frequency as well as the AoP, we show that the MDP problem \eqref{27} can be decomposed into two subproblems.  In the next section we will first solve the two subproblems for \eqref{27}, and then we propose an algorithm to obtain the optimal policy for the original CMDP problem \eqref{18}.

\section{Optimal policy for the CMDP problem}
In this section, we first propose a policy iteration algorithm to derive the optimal policy $\pi^{\lambda}$ for Subproblem 1. After that, we apply the Robbins-Monro algorithm to derive the optimal Lagrangian multiplier $\lambda^{*}$ for Subproblem 2. Finally, we propose an algorithm to derive the optimal policy for the original CMDP problem \eqref{18}.

\textbf{Solving Subproblem 1.} When given $\lambda$, problem \eqref{27} is a Markov decision process with an average reward criterion, which has been studied in many excellent works, e.g., \cite{puterman2014markov} and \cite{bertsekas1995dynamic}. We restrict the stationary policy $\pi$ to the stationary deterministic policy  $\pi_{sd}$.
%stationary deterministic policy, which simplifies the policy space and always guarantees the existence of the optimal policy of the MDP problem \eqref{27}. 
A stationary deterministic policy $\pi_{sd}$ maps each state to a single action. That is, given a state $A_i \in \mathcal{A}$, the output of policy $\pi_{sd}(A_i)$ is a single action, not a probability distribution over the action space. The stationary deterministic policy simplifies the state space and guarantees the existence of the optimal policy to the MDP problem \eqref{27}. 

Applying a stationary deterministic policy $\pi_{sd}$ to a controlled Markov process yields a Markov process with stationary transition probability matrix $\mathbf{P}_{\pi_{sd}}$, where the element $P_{\pi_{sd}}(A_i,A_j)$ is the state transition probability from $A_i$ to $A_j$ under policy $\pi_{sd}$ \cite{ross2014introduction}. Given policy $\pi_{sd}$, we also have a reward vector $\mathbf{r}_{\pi_{sd}} \in \mathbb{R}^{|\mathcal{A}|}$, where the element $r_{\pi_{sd}}(A_i)$ is the immediate reward $L^{\lambda}(A_i,\pi_{sd}(A_i))$ at state $A_i$ with the chosen action $\pi_{sd}(A_i)$. A gain vector $\mathbf{g}_{\pi_{sd}} \in \mathbb{R}^{|\mathcal{A}|}$ is an average reward vector, whose element $g_{\pi_{sd}}(A_i)$ is the average reward when starting at the initial state $A_i$

\begin{equation}
	g_{\pi_{sd}}(A_i) = \limsup_{n \rightarrow \infty}\frac{1}{n}\mathbb{E}_{\pi_{sd}}\left[\sum_{i=1}^{n}L^{\lambda}(A_i, \pi_{sd}(A_i))\right].
\end{equation}

Moreover, when given a $\lambda$, the MDP problem \eqref{27} has following Bellman optimality equation: 

\begin{equation}
	\overline{L}^{\lambda} + b_{\pi_{sd}}(A_i) = \min_{\bm{w}\in \mathcal{W}}\Bigl\{r_{\pi_{sd}}(A_i)+\sum_{A_j\in \mathcal{A}}P(A_j|A_i,\bm{w})b_{\pi_{sd}}(A_j)\Bigl\},
\end{equation}
where $P(A_j|A_i,\bm{w})$ is the probability from state $A_i$ to $A_j$ under the policy $\pi(A_i)=\bm{w}$, and the bias vector $\mathbf{b}_{\pi_{sd}} \in \mathbb{R}^{|\mathcal{A}|}$ is the expected total difference between the immediate reward and the average reward \cite{puterman2014markov}. Therefore, the optimal policy $\pi^{\lambda}$ can be obtained by:
\begin{equation}
	\pi^{\lambda}(A_i) = \arg \min_{\bm{w}\in \mathcal{W}}\Bigl\{r_{\pi_{sd}}(A_i)+\sum_{A_j\in \mathcal{A}}P(A_j|A_i,\bm{w})b_{\pi_{sd}}(A_j)\Bigl\}. \label{32}
\end{equation}

We propose the policy iteration algorithm to solve \eqref{32}, as shown in Algorithm 1. The key idea of Algorithm 1 is to iteratively perform policy evaluation and policy improvement to drive the update dynamics to converge to the optimal policy in \eqref{32}.   

\begin{algorithm}[t] 
	\caption{The policy iteration algorithm} 
	\begin{algorithmic}[1]
		\REQUIRE  
		Lagrangian multiplier $\lambda$;
		\ENSURE 
		The optimal policy $\pi_{sd}^{\lambda}$ of \eqref{27} when given a $\lambda$;
		\STATE Set $n=0$ and select an arbitrary stationary deterministic policy $\pi_{sd,0} \in \pi_{sd}$.
		\STATE (Policy evaluation) Obtain the average reward vector $\mathbf{g}_{\pi_{sd,n}}$, the bias vector $\mathbf{b}_{\pi_{sd,n}}$, and an auxiliary vector $\mathbf{\mu}_{\pi_{sd,n}}$ by solving a set of linear equations for $(\mathbf{g}_{\pi_{sd,n}},\mathbf{b}_{\pi_{sd,n}},\mathbf{\mu}_{\pi_{sd,n}})$ as follows:
		\begin{align}
		(I-P_{\pi_{sd,n}})\mathbf{g}_{\pi_{sd,n}}&=0,\\
		\mathbf{g}_{\pi_{sd,n}}+(I-P_{\pi_{sd,n}})\mathbf{b}_{\pi_{sd,n}}&=r_{\pi_{sd,n}},\\
		\mathbf{b}_{\pi_{sd,n}}+(I-P_{\pi_{sd,n}})\mathbf{\mu}_{\pi_{sd,n}}&=0,
		\end{align}
		where $I$ is a diagonal matrix with all one value and the same dimension as $P_{\pi_{sd,n}}$. $P_{\pi_{sd,n}}$and $r_{\pi_{sd,n}}$ are known when given $\pi_{sd,n}$.
		\STATE (Policy improvement)\\
		For each state $A_i \in \mathcal{A}$, choose $\pi_{sd,n+1} \in \pi_{sd}$ to satisfy 
		\begin{equation}
		\begin{aligned}
		\pi_{sd,n+1}(A_i) \in \arg \min_{\bm{w}\in \mathcal{W}}&\Bigl\{r_{\pi_{sd,n}}(A_i)+ \\ &\sum_{A_j\in \mathcal{A}}P(A_j|A_i,\bm{w})b_{\pi_{sd,n}}(A_j)\Bigl\}.
		\end{aligned}
		\end{equation}
			 
		setting $\pi_{sd,n+1}(A_i)=\pi_{sd,n}(A_i)$. \\
		\STATE If $\pi_{sd,n+1}=\pi_{sd,n}$, stop and set $\pi_{sd}^{\lambda}=\pi_{sd,n}$. Otherwise, increment $n$ by 1 and return to step 2.
	\end{algorithmic}
\end{algorithm}

The linear equations (33) and (34) can uniquely determine the gain vector $\mathbf{g}_{\pi_{sd}}$. However, as for $\mathbf{b}_{\pi_{sd}}$, the class of $\mathbf{b}_{\pi_{sd}} + k\bm{e}$, where $k$ is an arbitrary constant and $\bm{e}$ is an all one vector with the same dimension as $\mathbf{b}_{\pi_{sd}}$, all satisfy the linear equations (33) and (34). Therefore, an auxiliary vector $\mathbf{\mu}_{\pi_{sd,n}}$ and an additional equation (35) are introduced to determine $\mathbf{b}_{\pi_{sd}}$. Note that, in each iteration, the policy evaluation needs to solve a linear program with $3|\mathcal{A}|$ variables, and the  policy improvement needs conduct at most $|\mathcal{A}||\mathcal{W}|$ comparisons. The convergence of Algorithm 1 to the optimal policy of problem (26) can be shown by following the similar proof procedures in \renewcommand\citeright{, Sec. 8]}\cite{puterman2014markov}\renewcommand\citeright{]} and hence is omitted here for brevity.
\begin{comment}
	The following theorem guarantees that the policy iteration Algorithm 1 can convergence to the optimal policy of problem \eqref{27}.
	
	\begin{mythm}
	For a finite state and action space unichain Markov decision processes \eqref{27}, the policy iteration Algorithm 1 can convergence to the optimal policy.
	\end{mythm}
	
	\begin{proof}
	The detailed proof of this theorem can be obtained in \renewcommand\citeright{, Sec. 8]}\cite{puterman2014markov}\renewcommand\citeright{]}. Since this proof is quite lengthy, we omitted it for brevity.  
	\end{proof}
\end{comment}

\textbf{Solving Subproblem 2.} Since the minimum sampling duration  $\overline{T}(\pi^{\lambda}_{sd})$ is non-decreasing in the Lagrangian multiplier $\lambda$ according to Lemma 1, we adopt the two time-scale stochastic approximation based Robbins-Monro algorithm \cite{robbins1951stochastic} to solve Subproblem 2, as shown in Algorithm 2. Specifically, at the small time scale we solve the optimal policy for the MDP with a given Lagrange multiplier $\lambda^{k}$ (e.g., step 4 and 5), and at the large time scale we update the Lagrange multiplier according to

\begin{equation}
	\lambda^{k+1} = \lambda^k + \frac{1}{k}\left(T_{min}-\overline{T}(\pi^{\lambda}_{sd})\right), \label{35}
\end{equation}
(e.g., step 6 and 7). The sequence of Lagrange multipliers $(\lambda^1,\lambda^2,\ldots)$ derived by Algorithm 2 converges to $\lambda^{*}$ following the two time-scale stochastic approximation analysis in \cite{robbins1951stochastic}. 

\begin{algorithm}[t] 
	\caption{Lagrangian transformation algorithm for the CMDP problem \eqref{18}.} 
	\begin{algorithmic}[1]
		\REQUIRE  
		Stop criterion $C_{stop}$;
		\ENSURE 
		The policy $\pi_{sd}^{\lambda^{*}}$ of \eqref{18};
		\STATE \textbf{Initialization:}
		\STATE Initialized $\lambda$ with a sufficiently small number (e.g., $\lambda^1=0$) and $k=1$.
		\STATE \textbf{End initialization}
		\STATE \textbf{Repeat} transform the CMDP problem \eqref{18} to the MDP problem \eqref{27} when given a $\lambda^k$.
		\STATE \qquad Obtain the optimal policy $\pi_{sd}^{\lambda^k}$ for problem \eqref{27} using Algorithm 1.
		\STATE \qquad Update the Lagrange multiplier $\lambda$ according to \eqref{35}.	
		\STATE \qquad Increase $k$ by 1.
		\STATE \textbf{Until} some stop criterion are satisfied.
	\end{algorithmic}
\end{algorithm}

There are several possible stop criterion in Algorithm 2, for example, the difference between $|\lambda^{k+1}-\lambda^k|$ or $T_{min}-\overline{T}(\pi^{\lambda}_{sd})$ being small enough (e.g., smaller than  $C_{stop}=10^{-4}$), or the number in iterations of Algorithm 2 exceeding a prespecified number (e.g., $K=10^3$). In practice, the optimal Lagrange multiplier $\lambda^{*}$ derived by Algorithm 2 can be close to but not precisely the one defined at \eqref{31}. 
\begin{comment}
Therefore, we have no guarantee that the optimal policy of the original CMDP problem \eqref{18}, $\pi^{*}$, is equals to the policy $\pi_{sd}^{\lambda^{*}}$ of problem \eqref{27} when given $\lambda^{*}$.
\end{comment}
Nevertheless, when the $\lambda^{*}$ is close to the value defined in \eqref{31}, we can further refine the optimal policy $\widetilde{\pi}^{*}$ for \eqref{18} as follows.

%% file: Algorithm2.tex
\textbf{Solving Problem \eqref{18}.} We integrate the perturbation based refinement framework to achieving the optimal policy for problem \eqref{18}. We introduce two perturbed Lagrange multipliers $\lambda_1$ and $\lambda_2$ by imposing some perturbation to $\lambda^{*}$%\cite{sennott_1993} \cite{ma1986estimation}
. Given $\lambda^{*}$ derived by Algorithm 2, we set
\begin{equation}
	\lambda_1 = \lambda^{*}+\delta,\quad \lambda_2 = \lambda^{*}-\delta, \label{38}
\end{equation}
where $\delta$ is a small enough perturbation parameter (e.g., $\delta=10^{-4}$). Lemma 1 shows that $\overline{T}^\lambda$ is monotone non-decreasing in $\lambda$,
\begin{comment}
\begin{equation*}
\lim_{\lambda \uparrow \lambda^{*}}\overline{T}^{\lambda} \le T_{min} \le \lim_{\lambda \downarrow \lambda^{*}}\overline{T}^{\lambda},
\end{equation*}
\end{comment}
and hence

\begin{equation}
\overline{T}^{\lambda_2} \leq T_{min} \leq \overline{T}^{\lambda_1}.	
\end{equation}
Then we refine the optimal policy $\widetilde{\pi}^{*}$ as a randomized mixture of two perturbed policies $\pi^{\lambda_1}$ and  $\pi^{\lambda_2}$ as

\begin{equation}
\widetilde{\pi}^{*} =q\pi^{\lambda_1}+ (1-q)\pi^{\lambda_2}, \label{39}
\end{equation}
where the randomization factor $q$ can be given as

\begin{equation}
	q = \frac{T_{min}-\overline{T}^{\lambda_2}}{\overline{T}^{\lambda_1}-\overline{T}^{\lambda_2}}.
\end{equation}
In this way, we will satisfy the condition in \eqref{31} due to the fact that
\begin{equation}
	q\overline{T}^{\lambda_1}+(1-q)\overline{T}^{\lambda_2}=T_{min}.
\end{equation}
We summarize the policy refining procedure in Algorithm 3.
\begin{algorithm}[t]
	\caption{Optimal policy refining for CMDP problem \eqref{18}.}
	\begin{algorithmic}[1]
		\REQUIRE
		Stop criterion $C_{stop}$ and the perturbation value $\delta$;
		\ENSURE
		The optimal policy $\widetilde{\pi}^{*}$ of \eqref{18};
		\STATE Obtain the Lagrangian multiplier $\lambda^{*}$ using  Algorithm 2.
		\STATE Obtain $\lambda_1$ and $\lambda_2$ according to \eqref{38}.
		\STATE Obtain the policy $\pi^{\lambda_1}$ and $\pi^{\lambda_2}$ using Algorithm 1.
		\STATE Obtain the optimal policy $\widetilde{\pi}^{*}$ for the CMDP problem \eqref{18} according to \eqref{39}.
	\end{algorithmic}
\end{algorithm}

In Algorithm 3, when running the Algorithm 2 to obtain the optimal Lagrangian multiplier $\lambda^{*}$ in step 1, it takes a long time to converge due to the low convergence rate of the stochastic approximation technique in \eqref{35}. Since the step size $\frac{1}{k}$ is still large when $\lambda^{k}$ is near $\lambda^{*}$ after a few iterations, it would take a long time for $\frac{1}{k}$ to get small enough. Therefore, we improve Algorithm 2 by introducing a modified step size $\epsilon \times \frac{1}{k}$, where $\epsilon$ is small value (e.g., $10^{-3}$), and update $\lambda$ as:
\begin{equation}
	\lambda^{k+1} = \lambda^k + \epsilon \times\frac{1}{k}(T_{min}-\overline{T}(\pi^{\lambda}_{sd})). \label{41}
\end{equation}
\begin{figure}[t]
	\centering
	\subfigure[update $\lambda$ using \eqref{35}]{
		\begin{minipage}[t]{0.48\columnwidth}
			\centering
			\includegraphics[scale=0.18]{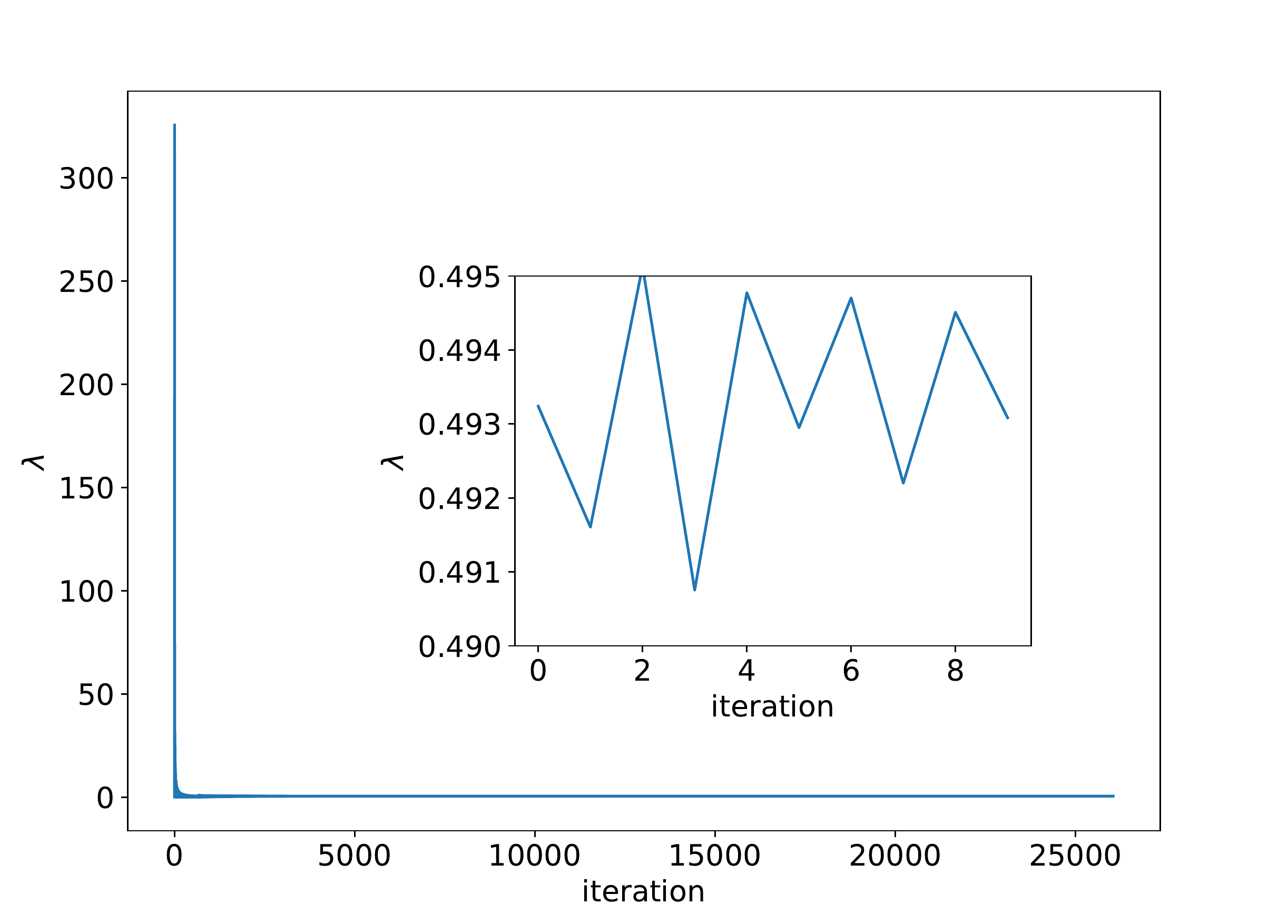}			
	\end{minipage}}
	\subfigure[update $\lambda$ using \eqref{41}]{
		\begin{minipage}[t]{0.48\columnwidth}
			\centering
			\includegraphics[scale=0.18]{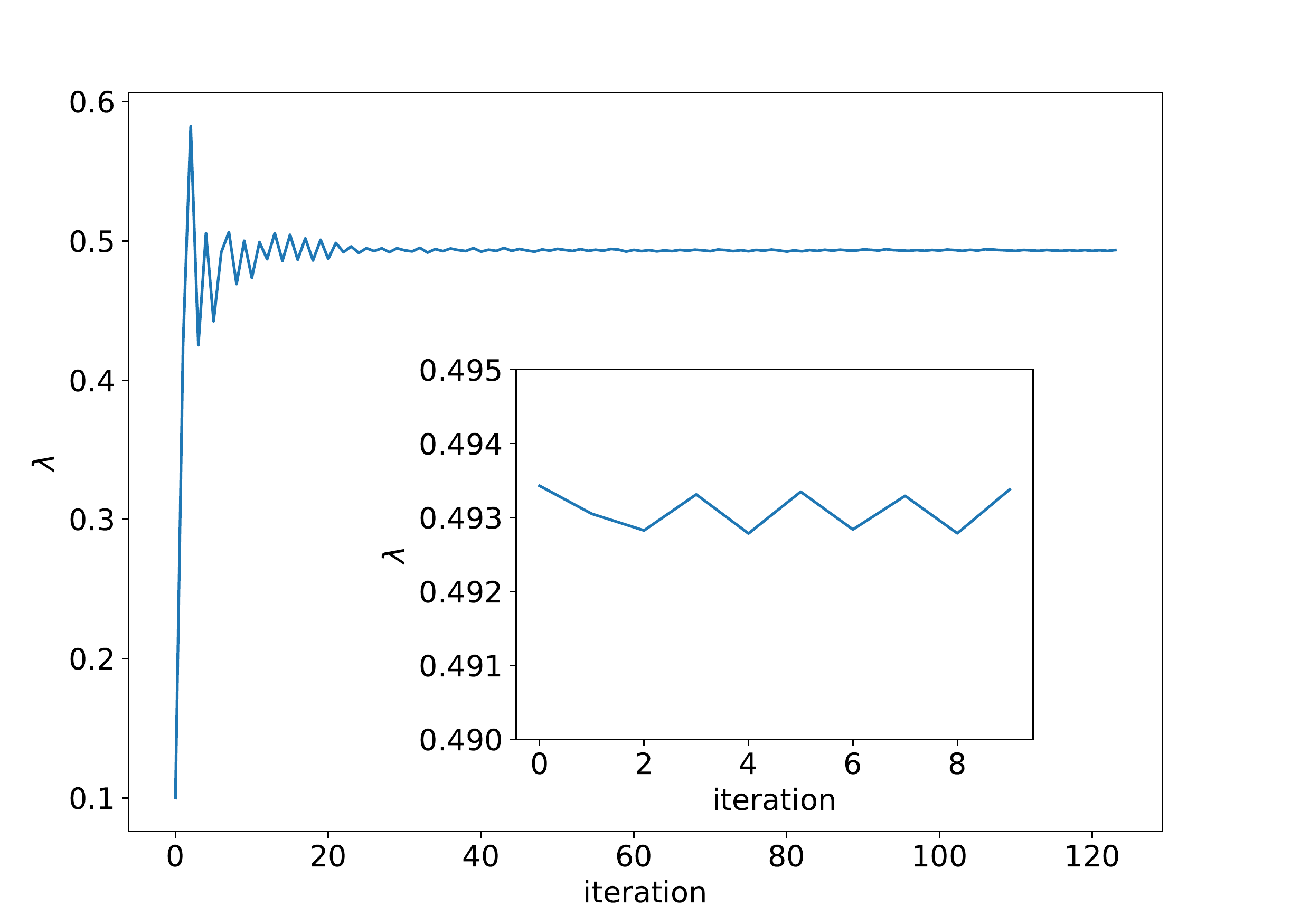}
	\end{minipage}}
\caption{value of $\lambda$ under different update step sizes.}
\end{figure}
As shown in Fig. 3(a), when updating $\lambda$ using $\eqref{35}$, it takes a long time to converge to the optimal Lagrangian multiplier $\lambda^{*}$ (e.g., more than 25000 iterations when the stop criterion here is $|\lambda^{k+1}-\lambda^{k}| \le 10^{-4}$). As shown in Fig. 3(b), the new updating rule \eqref{41} tremendously reduces the number of iterations (e.g., 120 iterations). Furthermore, the small figures in Fig. 3(a) and 3(b) show the last ten iterations of \eqref{35} and \eqref{41}. We can see that using \eqref{41} converges more close to $\lambda^{*}$.

%% file: Simulation.tex
\section{Performance evaluation }
In this section, we evaluate the performances of our proposed algorithms via extensive simulations.

\subsection{Simulation Setup}
As mentioned in Sec. III, we use $(l, c)$ to characterize the status update for an IoT computation-intensive application, where $l$ is the input data size and $c$ indicates the required CPU cycles. We also assume that all status update packets are of identical pair. Specially, we consider the face recognition application in \cite{6249269}, where the data size for the computation offloading $l=500$ KB and the total number of CPU cycles $c=1000$ Megacycles. In terms of computing resources, we assume that the CPU capability of edge and local server to be $f_e=20$ GHz and $f_l=1$ GHz \cite{tran2018joint}.

As for edge offloading, we assume that the wireless channel bandwidth $W=20$ MHz, and the distance between sensor and edge server $d=0.1$ km. The transmission power of sensor is $p=20$ dBm and the mean background noise $\sigma^2=-100$ dBm \cite{rappaport1996wireless}. We assume that the wireless channel state process is a Markov chain. Following the equal-probability SNR partitioning method \cite{zhang1999finite}, we can model the channel by three-state Markov chain, i.e., $\Gamma=\gamma_0,\gamma_1,\gamma_2$, with approximately the transition probability matrix
\begin{equation}
\mathbf{P}^{ch}=
	\begin{bmatrix}
		0.85 & 0.15 & 0 \\
		0.15 & 0.7  & 0.15 \\
		0    & 0.15 & 0.85
	\end{bmatrix}.
\end{equation}
Assume that if offloading is attempted, the transmission time defined in  \eqref{3} and \eqref{5} are given by $t^{tr}(\gamma_0)=500$ ms, $t^{tr}(\gamma_1)=1000$ ms, and $t^{tr}(\gamma_2)=2000$ ms. We summarize the main parameters of the simulation in Table I.

\begin{table}[tbp]
	\centering
	\caption{SIMULATION SETUP AND SYSTEM PARAMETERS}
	\begin{tabular}{lc}
		\toprule
		parameters	&  values\\
		\midrule
		input data size of each status update, $l$ & 500 KB\\
		number of CPU cycles of each status update, $c$  & 1000 Megacycles \\
		CPU cycle of edge server, $f_e$ & 20 GHz \\
		CPU cycle of local server, $f_l$ & 1 GHz \\
		wireless bandwidth between sensor and edge server, $W$ & 20 MHz \\
		distance between sensor and edge server, $d$ & 0.1 km \\
		transmission power of the sensor, $p$ & 20 dBm \\
		background noise, $\sigma^2$ & -100 dBm \\
		action set of waiting time, $Z$ & [0,200,...,800] ms \\
		minimum sampling duration, $T_{min}$ & 1200 ms \\
		perturbation parameter, $\delta$ & $3 \times 10^{-5}$ \\
		\bottomrule
	\end{tabular}
\end{table}

\subsection{Benchmarks}
In order to verify the performance of our proposed algorithms, we compare with the  following benchmarks:
\begin{enumerate}
	\item \textit{Always edge computing with zero waiting (AEZW)}: the sensor chooses to offload each status update to the edge server for further processing without waiting. That is, when the edge server completes computation of one status update, the sensor would sample an new status update immediately. However, this policy may not satisfy the sampling frequency constraint.
	\item \textit{Always edge computing with conservative waiting (AECW)}: the sensor chooses to offload each status update to the edge server with a conservative waiting. That is, when the edge server completes computation of one status update, based on current AoP $Y_i$ in the operator, the sensor choose to wait $\max\{T_{min}-Y_i, 0\}$ before sampling next status update.
	\item \textit{Always local computing with conservative waiting (ALCW)}: the sensor chooses to computer each status update at the local server with a conservative waiting. Since the local CPU cycle $f_l$ and total computation cycles $c$ are constants, the AoP $Y_i$ is also a constants when the local server completes computation, then the sensor choose to wait $T_{min}-Y_i$.
\end{enumerate}
\begin{figure}[t]
	\centering
	\includegraphics[scale=0.6]{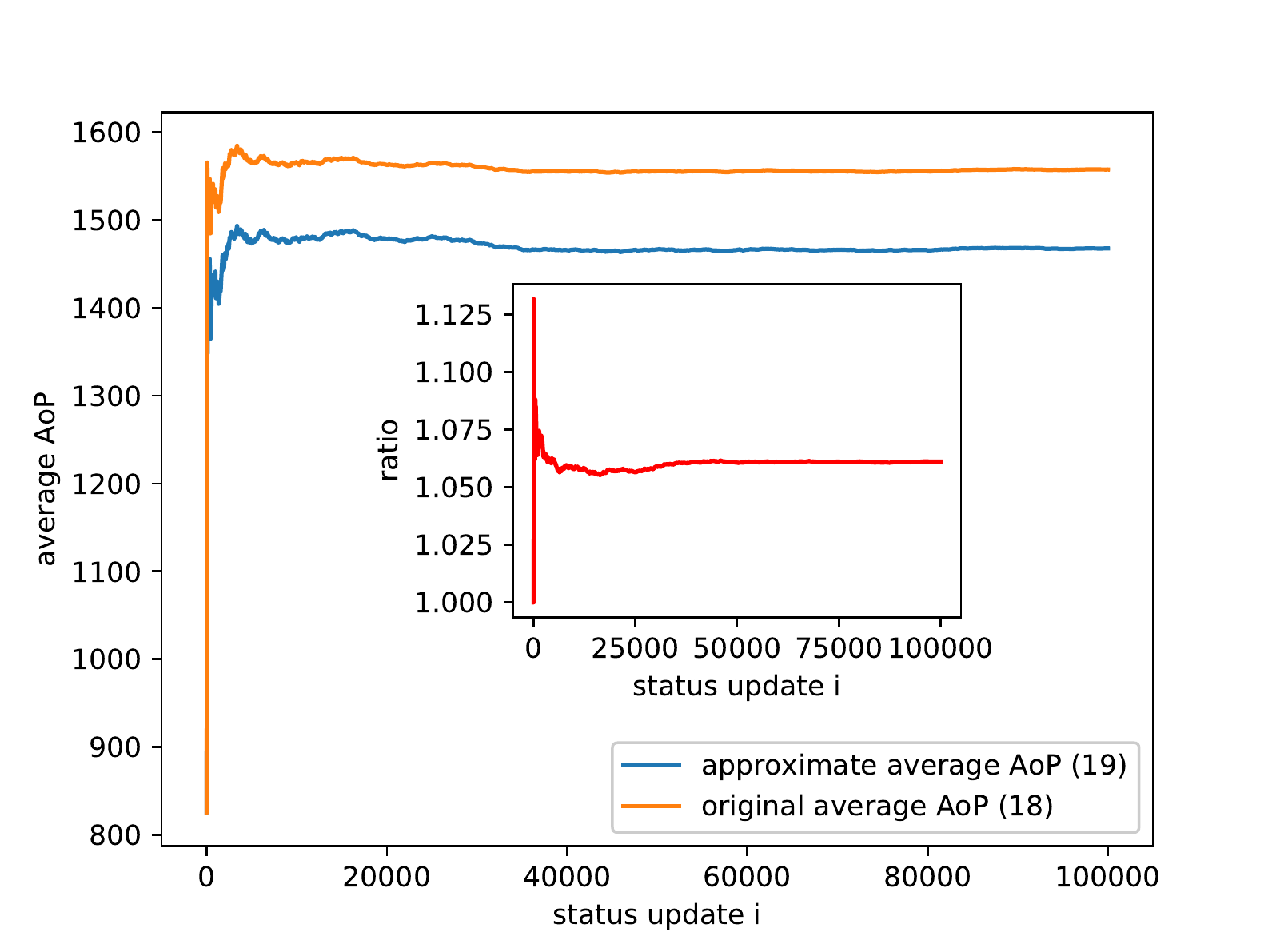}
	\caption{the average AoP of the original calculation \eqref{15} and the approximate calculation \eqref{19}.}
\end{figure}

\subsection{Policy structures of proposed algorithms}
We first compare the average AoP performance of the original problem \eqref{17} and the approximated problem \eqref{18}, and verify the optimal policy structure of the CMDP problem \eqref{18}.

As shown in Fig. 4, we conduct the simulation of $10^5$ status updates while using the optimal policy $\widetilde{\pi}^{*}$ defined in \eqref{39}. The orange line depicts the average AoP of the original problem \eqref{17} while the blue line depicts the approximated problem \eqref{18}. As we can see, when the status update number increases, the average AoP of both \eqref{17} and \eqref{18} would become stable, and the average AoP of \eqref{17} is slightly lager than that of \eqref{18}. More precisely, the small figure in Fig. 4 depicts the average AoP ratio of \eqref{17} and \eqref{18}. We can see that the ratio is very close to 1 (with the value of 1.06). This shows that, instead of obtaining the optimal policy of the original problem \eqref{17}, which is intractable, we seek to obtain the optimal policy $\widetilde{\pi}^{*}$ of the approximated problem \eqref{18}, and the solution $\widetilde{\pi}^{*}$ of \eqref{18} is also a nice approximation of the original problem \eqref{17}.

We depict the optimal policy structure of the CMDP problem \eqref{18} in the Fig. 5. The  coordinate $(y,x)$ represents the current system state $A_i$, where $x$ axis $x=(1,\ldots,6)$ represents the combination of the current AoP $Y_i$ and the wireless channel state $X_i$, while the $y$ axis $y=(1,\ldots,20)$ represents the combination of the last AoP $Y_{i-1}$ and waiting time $Z_{i-1}$. The $z$ axis $z=(1,\ldots,9)$ represents the action $\bm{w}_i=\pi(A_i)$ at state $A_i$, where even numbers denote ``offloading", odd numbers denote ``local computing", and bigger numbers represent higher waiting time (e.g., $z=3$ represents local computing and the waiting time is 200 ms). As shown in Fig. 5(a), the waiting time $Z_i$ is a threshold structure function of $Y_{i-1}$ and $Z_{i-1}$. That is, the optimal policy chooses longer waiting time $Z_i$ when the sum of the last AoP $Y_{i-1}$ and waiting time $Z_{i-1}$ is large (e.g., when $x$ axis is fixed and the value of $y$ axis increases, the value of $z$ axis also increases).

Fig. 5(a) shows the optimal policy structure $\pi^{\lambda^{*}}$ of the Lagrangian multiplier $\lambda^{*}$, which is obtained by Algorithm 2. Fig. 5(b) and 5(c) show the optimal policy $\pi^{\lambda_1}$ and $\pi^{\lambda_2}$, which are obtained by Algorithm 3. As we can see, the policy $\pi^{\lambda^{*}}$ and $\pi^{\lambda_1}$ are exactly the same when introducing a small perturbation $\delta$ to $\lambda^{*}$ ($\lambda_1 = \lambda^{*}+\delta$). Besides, the policy $\pi^{\lambda_1}$ and $\pi^{\lambda_2}$ only differ at one state which is pointed out by the red arrow in  state (3, 7).
\begin{figure*}
	\centering
	\subfigure[optimal policy $\pi^{\lambda^{*}}$ when given $\lambda^{*}$.]{
		\begin{minipage}[t]{0.66\columnwidth}
			\centering
			\includegraphics[scale=0.35]{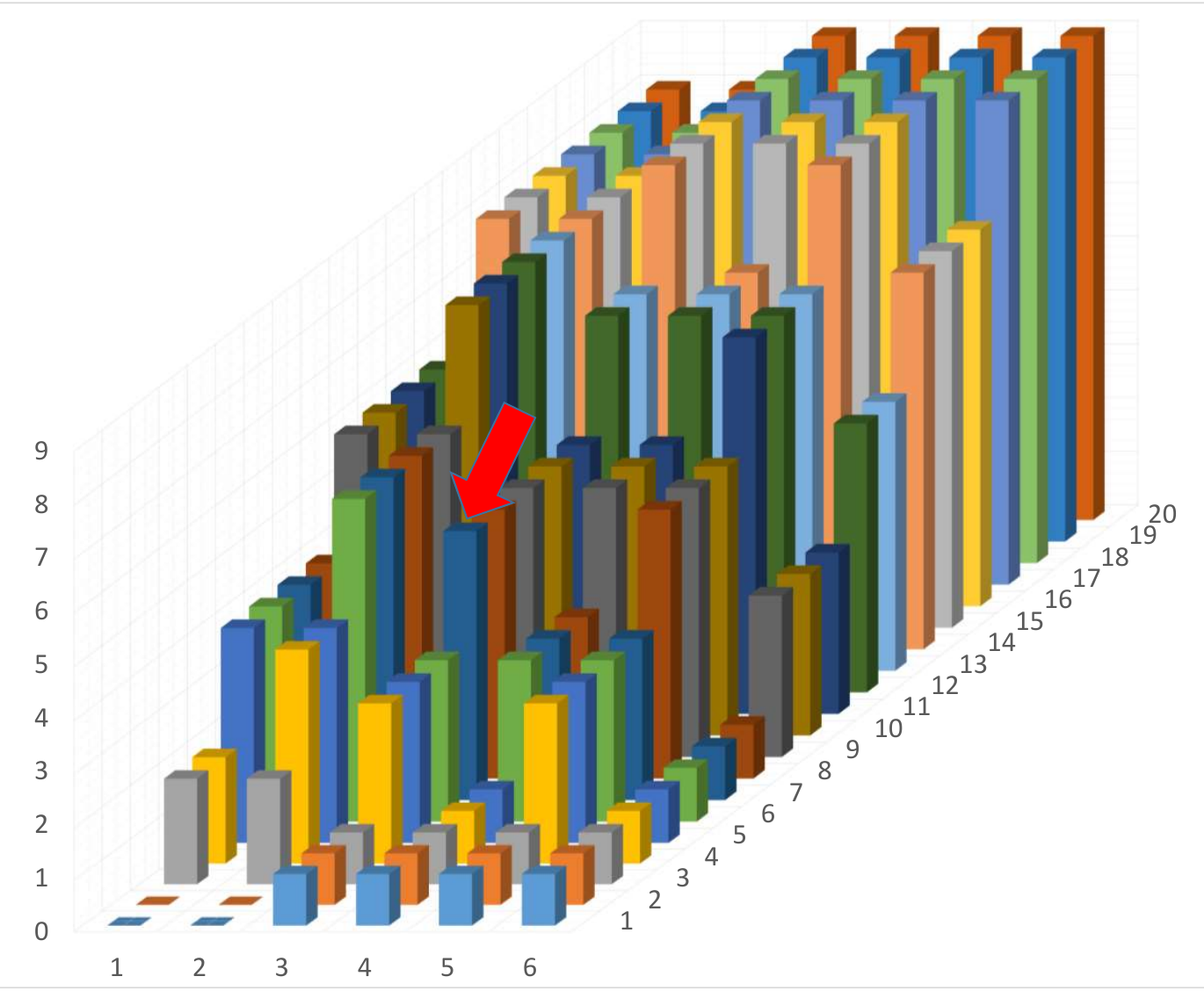}
	\end{minipage}}
	\subfigure[optimal policy $\pi^{\lambda_1}$ when given $\lambda_1$.]{
		\begin{minipage}[t]{0.66\columnwidth}
			\centering
			\includegraphics[scale=0.35]{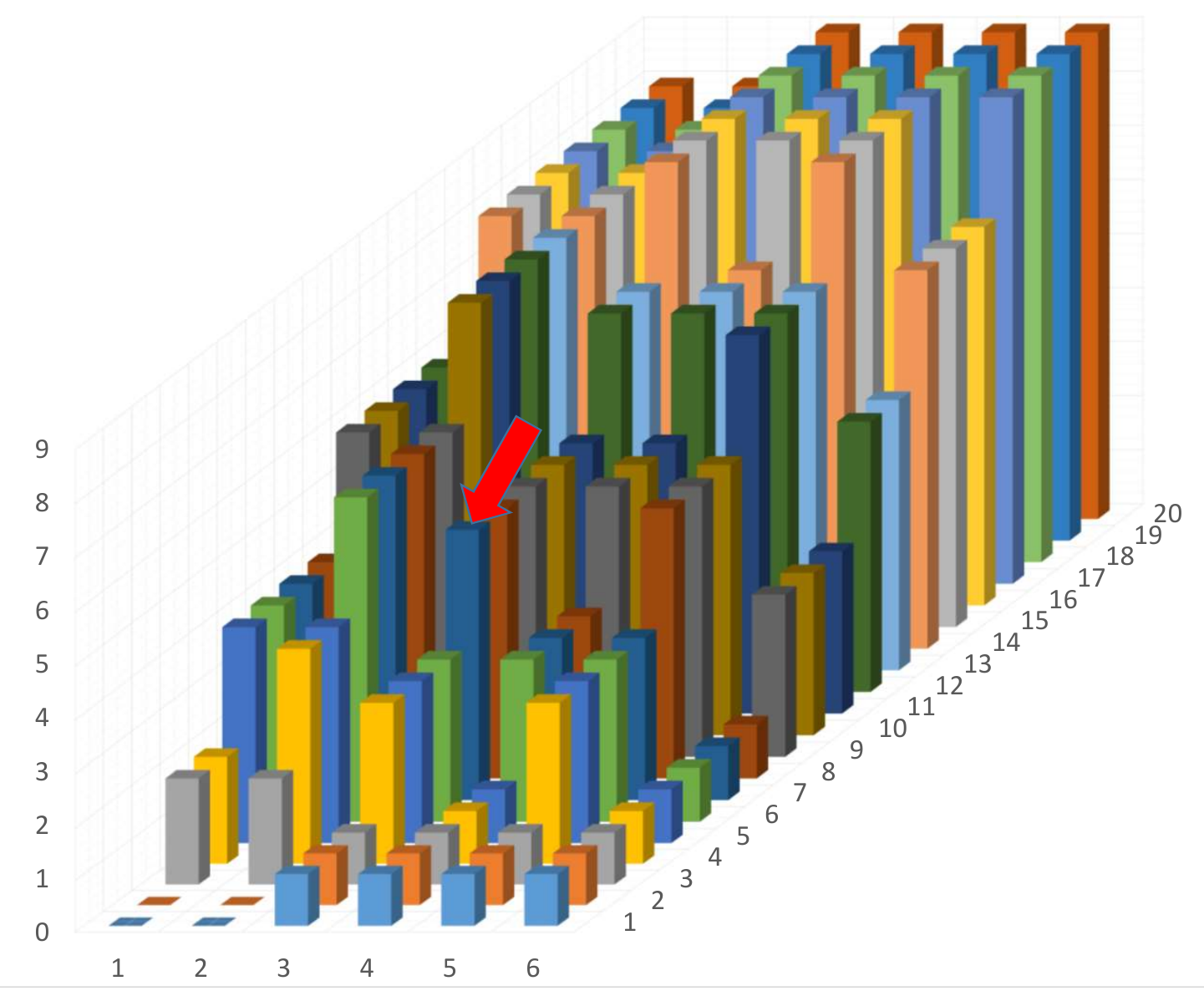}
	\end{minipage}}
	\subfigure[optimal policy $\pi^{\lambda_2}$ when given $\lambda_2$.]{
		\begin{minipage}[t]{0.66\columnwidth}
			\centering
			\includegraphics[scale=0.35]{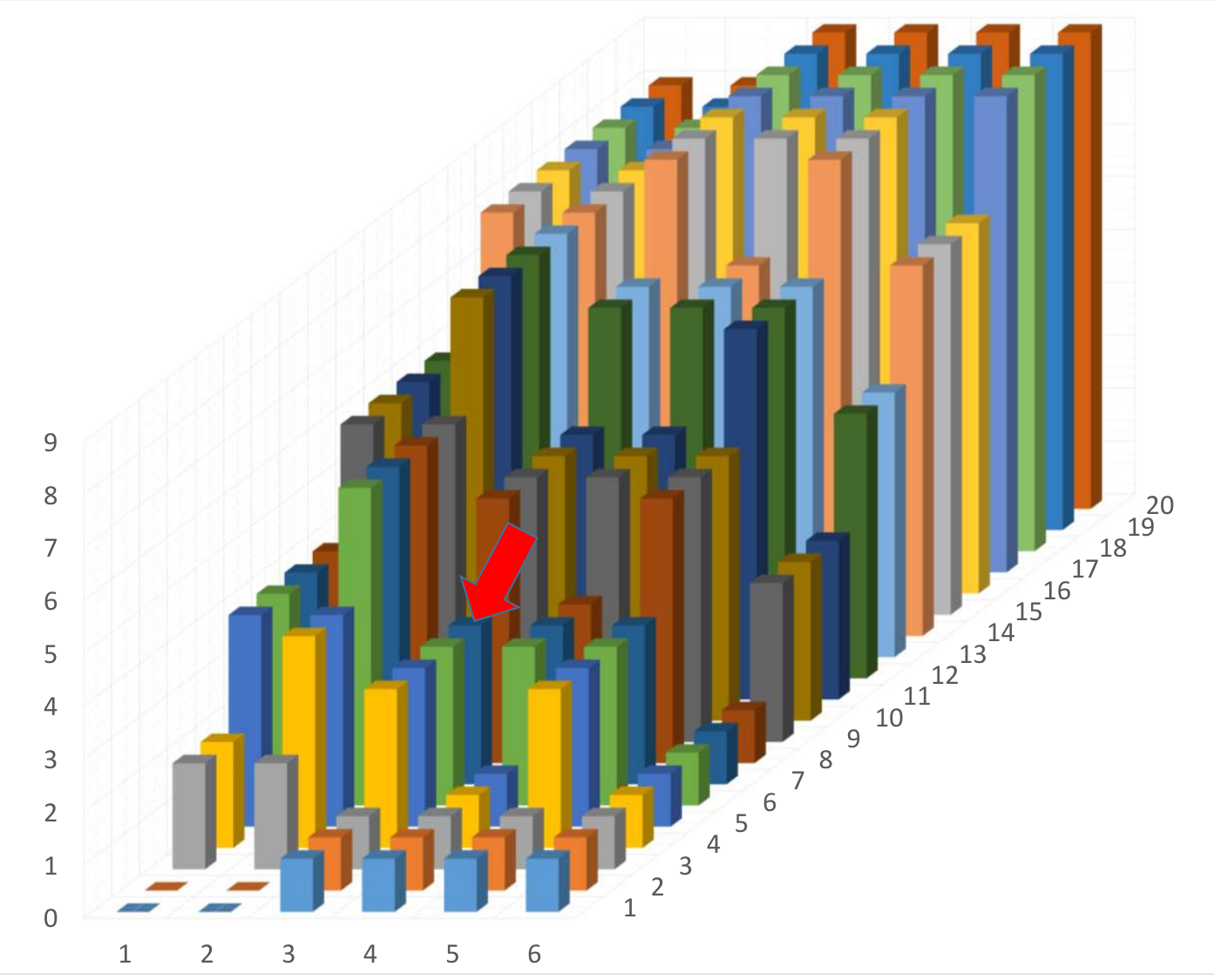}				
	\end{minipage}}
	\centering	
	\caption{different optimal policies $\pi$ when given different Lagrangian multipliers.}
\end{figure*}

\subsection{Performances among different benchmarks}
\begin{figure}[t]
	\centering
	\includegraphics[scale=0.6]{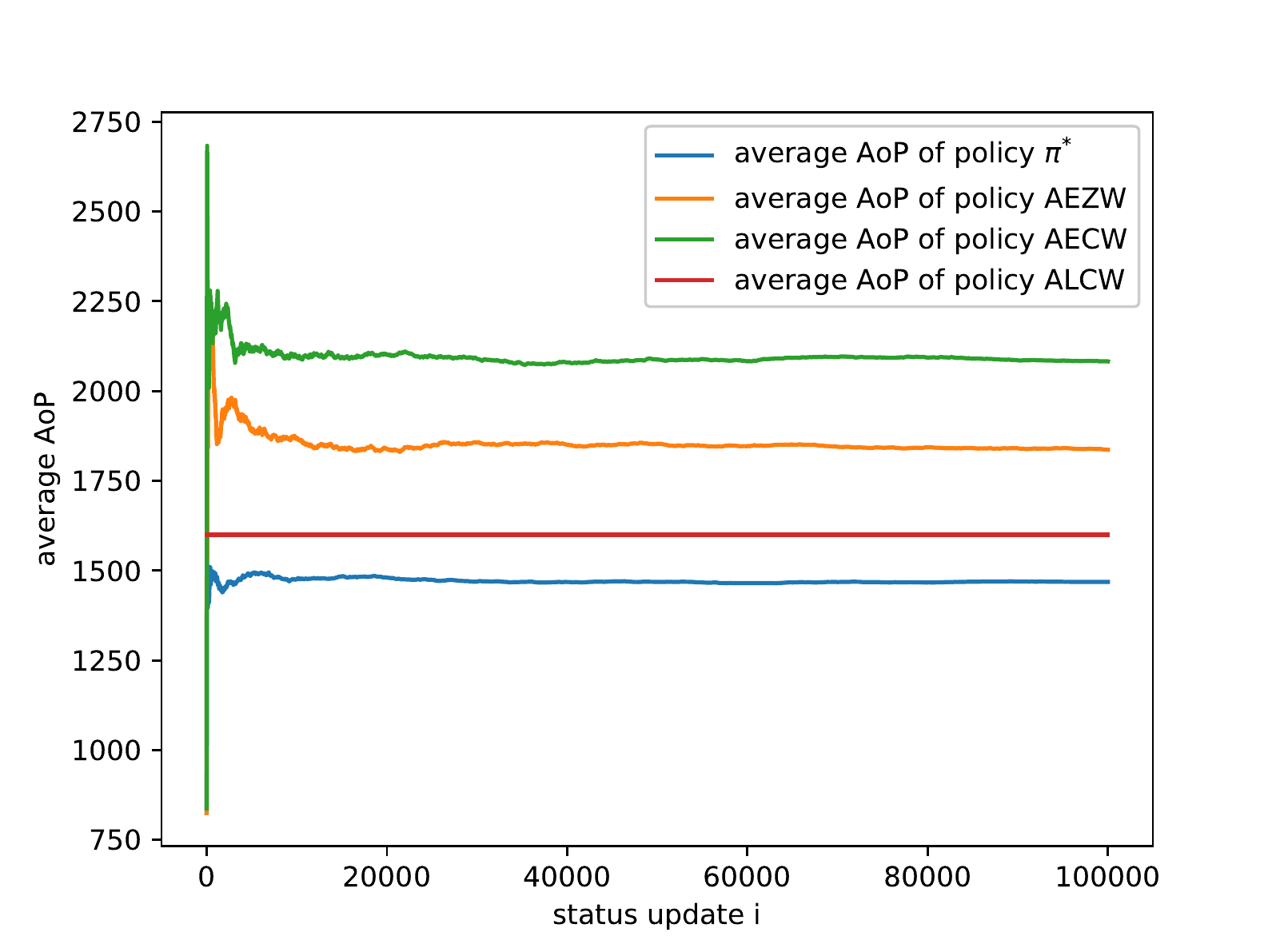}
	\caption{Average AoP performance among different policies.}
\end{figure}
\begin{figure*}[tbp]
	\centering
	\subfigure[average AoP of different transmission time]{
		\begin{minipage}[t]{1\columnwidth}
			\centering
			\includegraphics[scale=0.55]{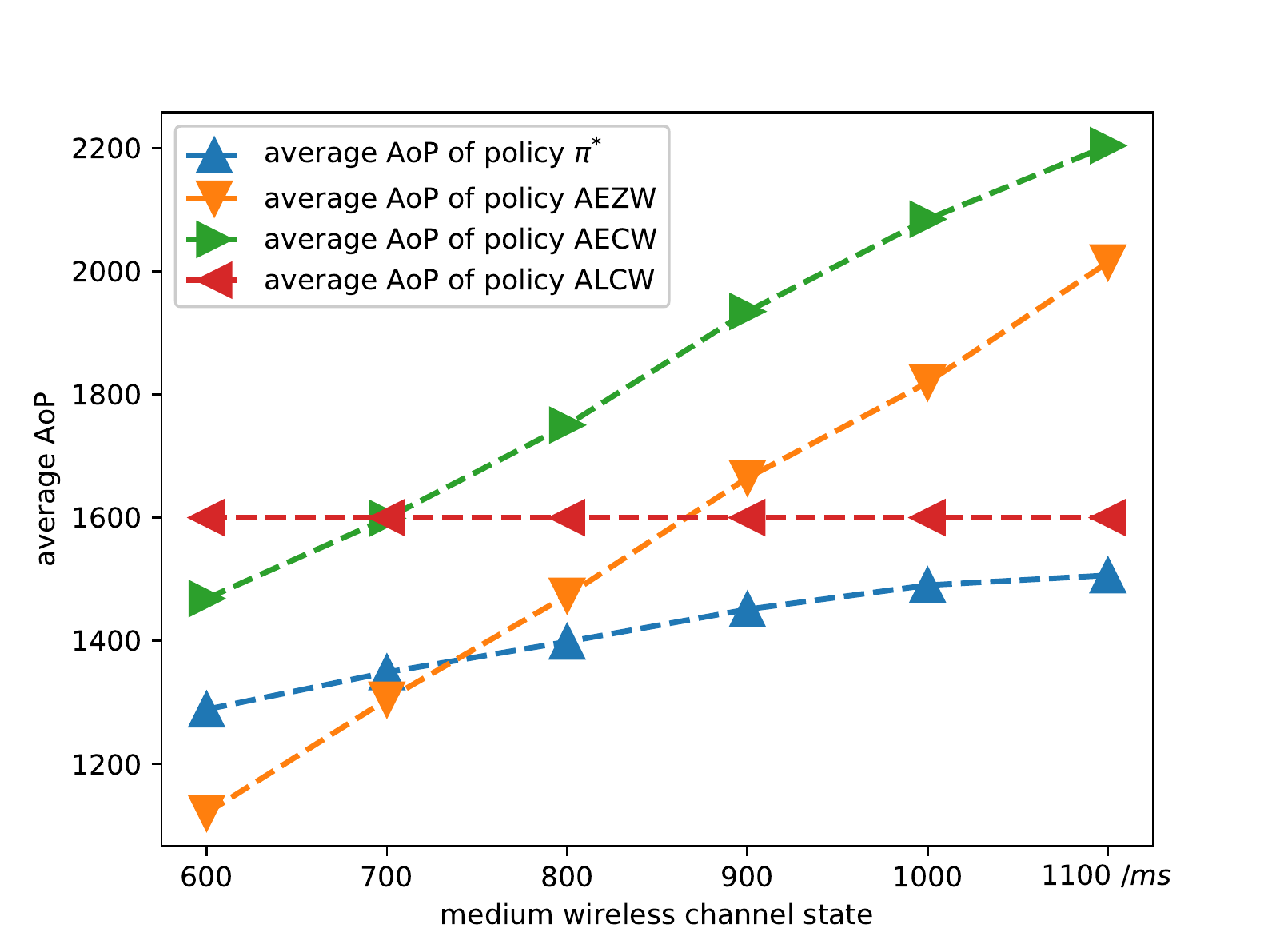}			
	\end{minipage}}
	\subfigure[average sampling time of different transmission time]{
		\begin{minipage}[t]{1\columnwidth}
			\centering
			\includegraphics[scale=0.55]{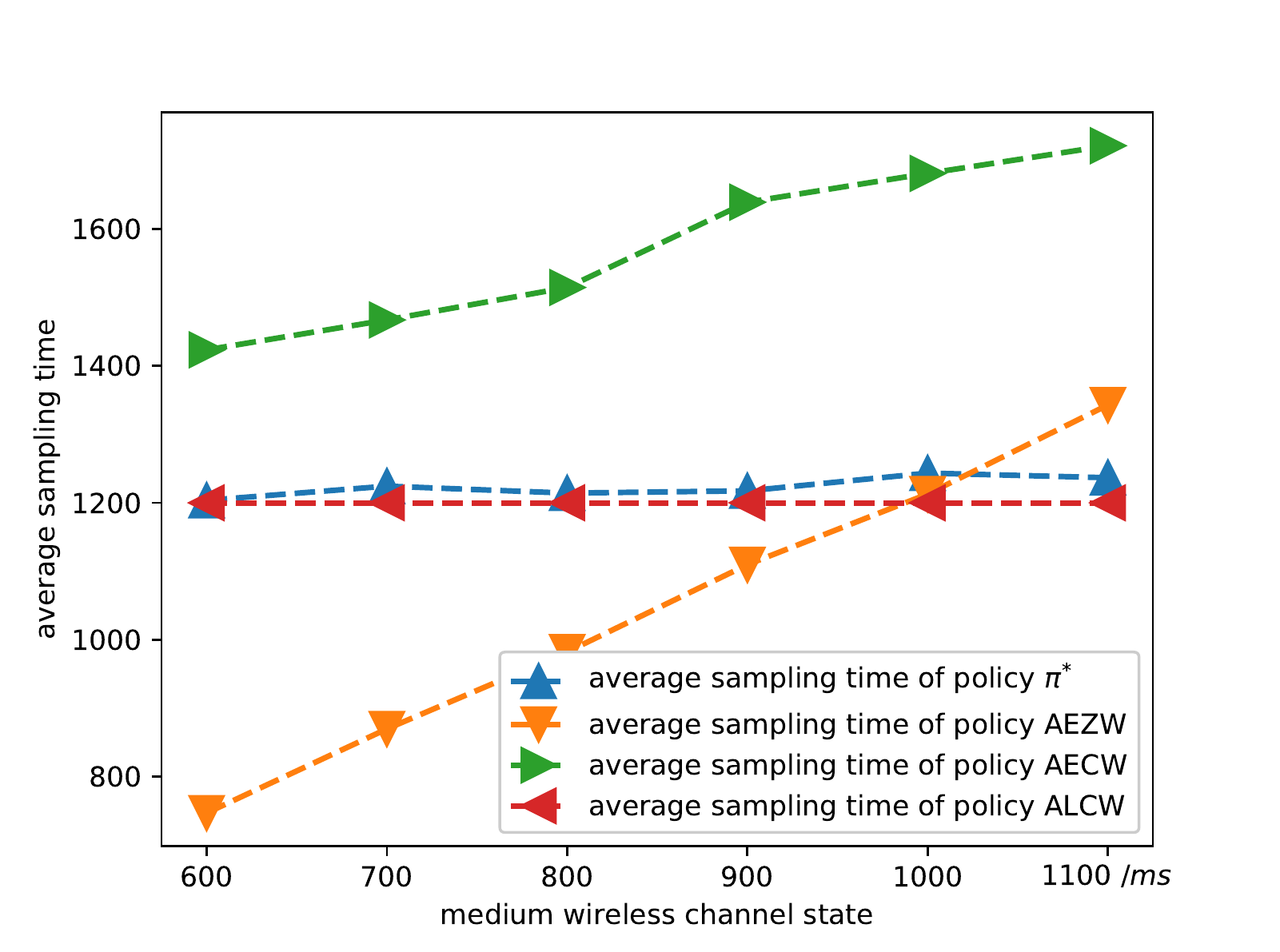}
	\end{minipage}}
	\caption{performance of four policy under different transmission time.}
\end{figure*}
We next conduct the simulations to compare average AoP performances among different benchmarks. As shown in Fig. 6, the optimal policy $\widetilde{\pi}^{*}$ achieves the minimum average AoP at around 1460 ms. The ALCW policy has a lower average AoP than AECW and AEZW, which is a constant of 1600 ms, and our proposed algorithms have an average AoP reduction at around 10\%. The reason of this reduction is that the optimal policy $\widetilde{\pi}^{*}$ would offload the status update to the edge server for further processing when the wireless channel state is good, and the powerful computing capacity of the edge server can shorten the processing time immensely, therefore, it results in a smaller average AoP. However, as shown in Fig. 6, the always offloading policies achieve a worse average AoP, at around 1840 ms and 2100 ms for AEZW and AECW, respectively, and our proposed algorithm achieves an AoP reduction at around 20\% and 30\%. The reason is that the average transmission time of offloading to the edge server is large in the original simulation setting. Although the processing time is small at edge server, the transmission time plays an critical role of AoP.

\subsection{The influence of wireless channel state}

In this subsection, we discuss the influence of wireless channel state for average AoP. Although the sensor can choose to offload to edge server to reduce the processing time, it would introduce additional transmission time. In this paper, we assume that the channel state is an Markov chain with three states. We can simply refer these three state to ``good", ``medium", and ``bad" channel state. We conduct the simulation with different transmission time of the medium channel state (e.g., $[600, 700, \ldots, 1100]$ ms)\footnote{the transmission time of good channel state is half of the medium state, and the transmission time of bad channel state is twice of the medium state.}. As shown in Fig. 7(a), when the transmission time increases, the average AoP of our proposed algorithm and the always offloading policies (AEZW and AECW) also increases. Besides, our algorithm has a much smaller increase rate,  because the optimal policy would choose to local computing when the wireless channel state is bad. When the transmission time less than 700 ms, the AEZW policy has a smaller average AoP than our proposed algorithm, however, as shown in Fig. 7(b), the average sampling time of AEZW is less than $T_{min}=1200$ ms, which violates the sampling frequency constraint \eqref{16}. Although the AECW and ALCW policies can always satisfy the constraint \eqref{16}, they result in a worse average AoP.

\subsection{The influence of computation demand}
\begin{figure}[t]
	\centering
	\includegraphics[scale=0.6]{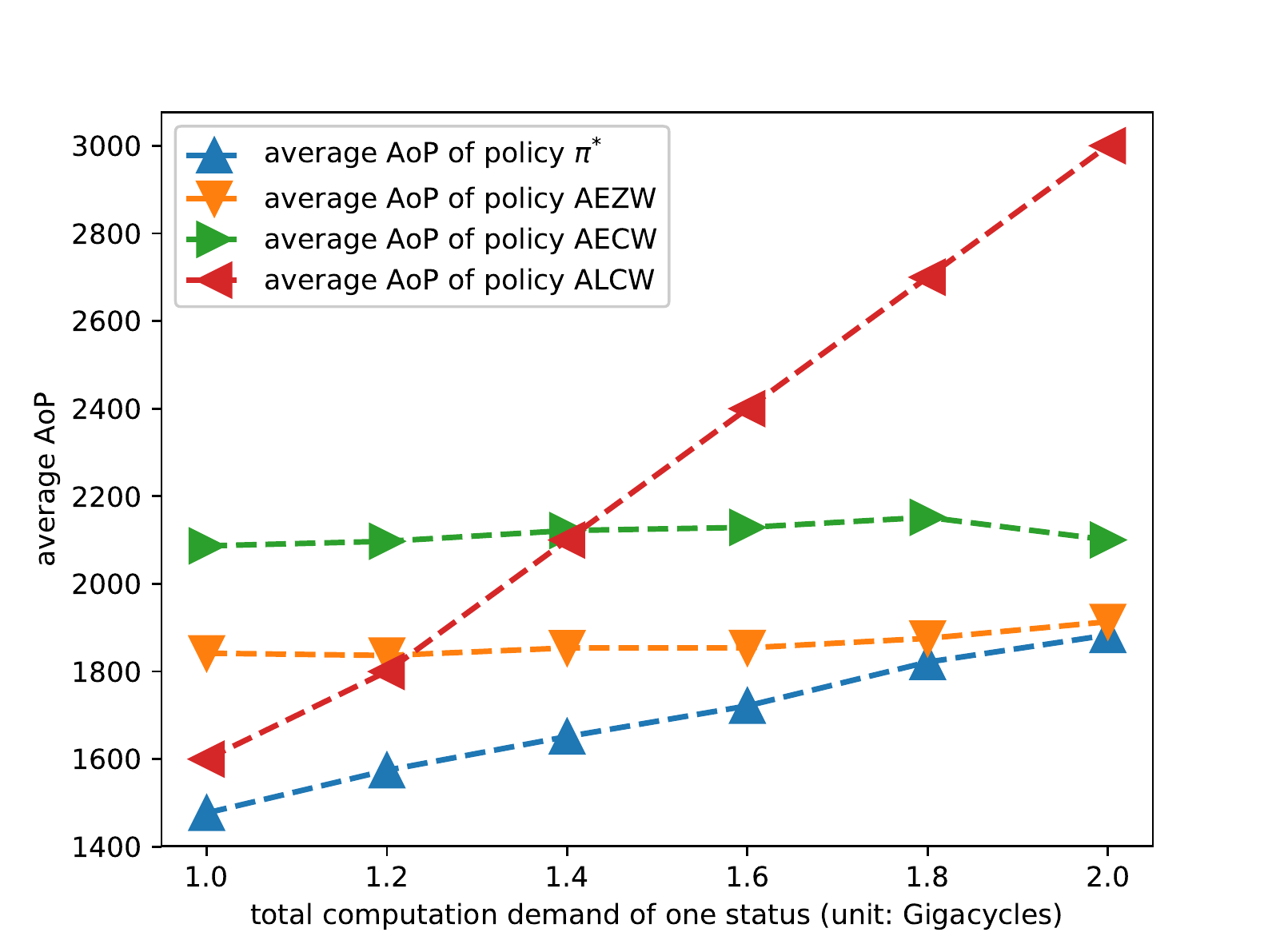}
	\caption{Average AoP performance under different computation demand.}
\end{figure}
In this subsection, we discuss the influence of computation demand for average AoP. We conduct the simulation of different computation demand of one status update (e.g., $[1.0, 1.2, \ldots, 2.0]$ Gigacycles) while the transmission time of the medium channel state is 1000 ms. As shown in Fig. 8, the average AoP of the ALCW policy increases dramatically when the computation demand increases from 1.0 to 2.0 Gigacycles due to the limited computation capacity of the local server. It takes much time to process a status update for computation-intensive application at the local server. In contrast, the average AoP of always offloading policies AEZW and AECW just has a slight increment since the edge server has a much larger computation capacity. We should note that, when the computation demand is 2.0 Gigacycles, the average AoP of our proposed algorithm equals to that of the AEZW policy. The reason is that when the computation demand is essentially large, the processing time would dominates the AoP, the proposed algorithm would choose to always offloading policy to reduce the processing time.

%% file: Conclusion.tex
\section{Conclusion}
In this paper, we aim to minimize the age-of-processing (AoP) of computation-intensive IoT application in a status monitoring and control system. Due to the limited resource of an IoT sensor, it can offload the status update to the edge server for processing. We focus on finding the optimal sampling and processing offloading policy to minimize the average AoP, which is formulated as a CMDP. We propose a Lagrangian transformation method to relax the CMDP problem into an unconstrained MDP problem, and derive the optimal policy when given the optimal Lagrangian multiplier of the MDP problem. Furthermore, by introducing a small perturbation value to the optimal Lagrangian multiplier of the MDP problem, we obtain the optimal policy of the original CMDP problem. The extensive simulation results verify the superior performance of our proposed algorithms. For the future direction, we are going to generalize our framework to the much more challenging scenarios with multiple IoT devices and edge servers.

%% file: AoP.bbl
\begin{thebibliography}{10}
\providecommand{\url}[1]{#1}
\csname url@samestyle\endcsname
\providecommand{\newblock}{\relax}
\providecommand{\bibinfo}[2]{#2}
\providecommand{\BIBentrySTDinterwordspacing}{\spaceskip=0pt\relax}
\providecommand{\BIBentryALTinterwordstretchfactor}{4}
\providecommand{\BIBentryALTinterwordspacing}{\spaceskip=\fontdimen2\font plus
\BIBentryALTinterwordstretchfactor\fontdimen3\font minus
  \fontdimen4\font\relax}
\providecommand{\BIBforeignlanguage}[2]{{%
\expandafter\ifx\csname l@#1\endcsname\relax
\typeout{** WARNING: IEEEtran.bst: No hyphenation pattern has been}%
\typeout{** loaded for the language `#1'. Using the pattern for}%
\typeout{** the default language instead.}%
\else
\language=\csname l@#1\endcsname
\fi
#2}}
\providecommand{\BIBdecl}{\relax}
\BIBdecl

\bibitem{atzori2010internet}
L.~Atzori, A.~Iera, and G.~Morabito, ``The internet of things: A survey,''
  \emph{Computer networks}, vol.~54, no.~15, pp. 2787--2805, 2010.

\bibitem{8793011}
T.~{Shreedhar}, S.~K. {Kaul}, and R.~D. {Yates}, ``An age control transport
  protocol for delivering fresh updates in the internet-of-things,'' in
  \emph{2019 IEEE 20th International Symposium on "A World of Wireless, Mobile
  and Multimedia Networks" (WoWMoM)}, June 2019, pp. 1--7.

\bibitem{DBLP:journals/corr/abs-1902-06149}
\BIBentryALTinterwordspacing
B.~Li and J.~Liu, ``Can we achieve fresh information with selfish users in
  mobile crowd-learning?'' \emph{CoRR}, vol. abs/1902.06149, 2019. [Online].
  Available: \url{http://arxiv.org/abs/1902.06149}
\BIBentrySTDinterwordspacing

\bibitem{5549870}
V.~{Terzija}, G.~{Valverde}, D.~{Cai}, P.~{Regulski}, V.~{Madani}, J.~{Fitch},
  S.~{Skok}, M.~M. {Begovic}, and A.~{Phadke}, ``Wide-area monitoring,
  protection, and control of future electric power networks,''
  \emph{Proceedings of the IEEE}, vol.~99, no.~1, pp. 80--93, Jan 2011.

\bibitem{8555643}
S.~{Zhang}, J.~{Li}, H.~{Luo}, J.~{Gao}, L.~{Zhao}, and X.~S. {Shen}, ``Towards
  fresh and low-latency content delivery in vehicular networks: An edge caching
  aspect,'' in \emph{2018 10th International Conference on Wireless
  Communications and Signal Processing (WCSP)}, Oct 2018, pp. 1--6.

\bibitem{kaul2011minimizing}
S.~Kaul, M.~Gruteser, V.~Rai, and J.~Kenney, ``Minimizing age of information in
  vehicular networks,'' in \emph{2011 8th Annual IEEE Communications Society
  Conference on Sensor, Mesh and Ad Hoc Communications and Networks}.\hskip 1em
  plus 0.5em minus 0.4em\relax IEEE, 2011, pp. 350--358.

\bibitem{NET-060}
\BIBentryALTinterwordspacing
A.~Kosta, N.~Pappas, and V.~Angelakis, ``Age of information: A new concept,
  metric, and tool,'' \emph{Foundations and Trends® in Networking}, vol.~12,
  no.~3, pp. 162--259, 2017. [Online]. Available:
  \url{http://dx.doi.org/10.1561/1300000060}
\BIBentrySTDinterwordspacing

\bibitem{moltafet2019age}
M.~Moltafet, M.~Leinonen, and M.~Codreanu, ``On the age of information in
  multi-source queueing models,'' 2019.

\bibitem{talak2019age}
R.~Talak and E.~Modiano, ``Age-delay tradeoffs in queueing systems,''
  \emph{arXiv preprint arXiv:1911.05601}, 2019.

\bibitem{akar2019finding}
N.~Akar, O.~Dogan, and E.~U. Atay, ``Finding the exact distribution of (peak)
  age of information for queues of ph/ph/1/1 and m/ph/1/2 type,'' 2019.

\bibitem{DBLP:journals/corr/abs-1906-12278}
\BIBentryALTinterwordspacing
J.~Xu and N.~Gautam, ``Towards assigning priorities in queues using age of
  information,'' \emph{CoRR}, vol. abs/1906.12278, 2019. [Online]. Available:
  \url{http://arxiv.org/abs/1906.12278}
\BIBentrySTDinterwordspacing

\bibitem{DBLP:journals/corr/abs-1901-08197}
\BIBentryALTinterwordspacing
M.~Wang, W.~Chen, and A.~Ephremides, ``Real-time reconstruction of counting
  process through queues,'' \emph{CoRR}, vol. abs/1901.08197, 2019. [Online].
  Available: \url{http://arxiv.org/abs/1901.08197}
\BIBentrySTDinterwordspacing

\bibitem{DBLP:journals/corr/abs-1905-13743}
\BIBentryALTinterwordspacing
A.~Soysal and S.~Ulukus, ``Age of information in {G/G/1/1} systems: Age
  expressions, bounds, special cases, and optimization,'' \emph{CoRR}, vol.
  abs/1905.13743, 2019. [Online]. Available:
  \url{http://arxiv.org/abs/1905.13743}
\BIBentrySTDinterwordspacing

\bibitem{DBLP:journals/corr/abs-1901-10463}
\BIBentryALTinterwordspacing
V.~Tripathi, R.~Talak, and E.~Modiano, ``Age of information for discrete time
  queues,'' \emph{CoRR}, vol. abs/1901.10463, 2019. [Online]. Available:
  \url{http://arxiv.org/abs/1901.10463}
\BIBentrySTDinterwordspacing

\bibitem{8445909}
H.~{Sac}, T.~{Bacinoglu}, E.~{Uysal-Biyikoglu}, and G.~{Durisi}, ``Age-optimal
  channel coding blocklength for an m/g/1 queue with harq,'' in \emph{2018 IEEE
  19th International Workshop on Signal Processing Advances in Wireless
  Communications (SPAWC)}, June 2018, pp. 1--5.

\bibitem{DBLP:journals/corr/abs-1806-09396}
\BIBentryALTinterwordspacing
R.~Devassy, G.~Durisi, G.~C. Ferrante, O.~Simeone, and E.~Uysal{-}Biyikoglu,
  ``Reliable transmission of short packets through queues and noisy channels
  under latency and peak-age violation guarantees,'' \emph{CoRR}, vol.
  abs/1806.09396, 2018. [Online]. Available:
  \url{http://arxiv.org/abs/1806.09396}
\BIBentrySTDinterwordspacing

\bibitem{DBLP:journals/corr/abs-1804-06139}
\BIBentryALTinterwordspacing
Y.~Inoue, H.~Masuyama, T.~Takine, and T.~Tanaka, ``A general formula for the
  stationary distribution of the age of information and its application to
  single-server queues,'' \emph{CoRR}, vol. abs/1804.06139, 2018. [Online].
  Available: \url{http://arxiv.org/abs/1804.06139}
\BIBentrySTDinterwordspacing

\bibitem{8406909}
J.~P. {Champati}, H.~{Al-Zubaidy}, and J.~{Gross}, ``Statistical guarantee
  optimization for age of information for the d/g/1 queue,'' in \emph{IEEE
  INFOCOM 2018 - IEEE Conference on Computer Communications Workshops (INFOCOM
  WKSHPS)}, April 2018, pp. 130--135.

\bibitem{DBLP:journals/corr/abs-1801-04068}
\BIBentryALTinterwordspacing
E.~Najm and E.~Telatar, ``Status updates in a multi-stream {M/G/1/1} preemptive
  queue,'' \emph{CoRR}, vol. abs/1801.04068, 2018. [Online]. Available:
  \url{http://arxiv.org/abs/1801.04068}
\BIBentrySTDinterwordspacing

\bibitem{DBLP:journals/corr/abs-1709-04956}
\BIBentryALTinterwordspacing
A.~M. Bedewy, Y.~Sun, and N.~B. Shroff, ``Minimizing the age of the information
  through queues,'' \emph{CoRR}, vol. abs/1709.04956, 2017. [Online].
  Available: \url{http://arxiv.org/abs/1709.04956}
\BIBentrySTDinterwordspacing

\bibitem{8006592}
Y.~{Inoue}, H.~{Masuyama}, T.~{Takine}, and T.~{Tanaka}, ``The stationary
  distribution of the age of information in fcfs single-server queues,'' in
  \emph{2017 IEEE International Symposium on Information Theory (ISIT)}, June
  2017, pp. 571--575.

\bibitem{DBLP:journals/corr/abs-1805-12586}
\BIBentryALTinterwordspacing
A.~Soysal and S.~Ulukus, ``Age of information in {G/G/1/1} systems,''
  \emph{CoRR}, vol. abs/1805.12586, 2018. [Online]. Available:
  \url{http://arxiv.org/abs/1805.12586}
\BIBentrySTDinterwordspacing

\bibitem{10.1287/opre.9.3.383}
\BIBentryALTinterwordspacing
J.~D.~C. Little, ``A proof for the queuing formula: L = $\lambda$w,''
  \emph{Oper. Res.}, vol.~9, no.~3, p. 383–387, Jun. 1961. [Online].
  Available: \url{https://doi.org/10.1287/opre.9.3.383}
\BIBentrySTDinterwordspacing

\bibitem{6195689}
S.~{Kaul}, R.~{Yates}, and M.~{Gruteser}, ``Real-time status: How often should
  one update?'' in \emph{2012 Proceedings IEEE INFOCOM}, March 2012, pp.
  2731--2735.

\bibitem{8815809}
R.~{Li}, Z.~{Zhou}, X.~{Chen}, and Q.~{Ling}, ``Resource price-aware offloading
  for edge-cloud collaboration: A two-timescale online control approach,''
  \emph{IEEE Transactions on Cloud Computing}, pp. 1--1, 2019.

\bibitem{8684949}
B.~{Barakat}, S.~{Keates}, I.~{Wassell}, and K.~{Arshad}, ``Is the zero-wait
  policy always optimum for information freshness (peak age) or throughput?''
  \emph{IEEE Communications Letters}, vol.~23, no.~6, pp. 987--990, June 2019.

\bibitem{7283009}
R.~D. {Yates}, ``Lazy is timely: Status updates by an energy harvesting
  source,'' in \emph{2015 IEEE International Symposium on Information Theory
  (ISIT)}, June 2015, pp. 3008--3012.

\bibitem{7524524}
Y.~{Sun}, E.~{Uysal-Biyikoglu}, R.~{Yates}, C.~E. {Koksal}, and N.~B. {Shroff},
  ``Update or wait: How to keep your data fresh,'' in \emph{IEEE INFOCOM 2016 -
  The 35th Annual IEEE International Conference on Computer Communications},
  April 2016, pp. 1--9.

\bibitem{6875100}
M.~{Costa}, M.~{Codreanu}, and A.~{Ephremides}, ``Age of information with
  packet management,'' in \emph{2014 IEEE International Symposium on
  Information Theory}, June 2014, pp. 1583--1587.

\bibitem{DBLP:journals/corr/abs-1907-03826}
\BIBentryALTinterwordspacing
G.~Stamatakis, N.~Pappas, and A.~Traganitis, ``Control of status updates for
  energy harvesting devices that monitor processes with alarms,'' \emph{CoRR},
  vol. abs/1907.03826, 2019. [Online]. Available:
  \url{http://arxiv.org/abs/1907.03826}
\BIBentrySTDinterwordspacing

\bibitem{ceran2019reinforcement}
E.~T. Ceran, D.~Gündüz, and A.~György, ``Reinforcement learning to minimize
  age of information with an energy harvesting sensor with harq and sensing
  cost,'' 2019.

\bibitem{Arafa_2018}
\BIBentryALTinterwordspacing
A.~Arafa, J.~Yang, and S.~Ulukus, ``Age-minimal online policies for energy
  harvesting sensors with random battery recharges,'' \emph{2018 IEEE
  International Conference on Communications (ICC)}, May 2018. [Online].
  Available: \url{http://dx.doi.org/10.1109/ICC.2018.8422086}
\BIBentrySTDinterwordspacing

\bibitem{DBLP:journals/corr/abs-1802-02129}
\BIBentryALTinterwordspacing
A.~Arafa, J.~Yang, S.~Ulukus, and H.~V. Poor, ``Age-minimal online policies for
  energy harvesting sensors with incremental battery recharges,'' \emph{CoRR},
  vol. abs/1802.02129, 2018. [Online]. Available:
  \url{http://arxiv.org/abs/1802.02129}
\BIBentrySTDinterwordspacing

\bibitem{8635855}
M.~{Bastopcu} and S.~{Ulukus}, ``Age of information with soft updates,'' in
  \emph{2018 56th Annual Allerton Conference on Communication, Control, and
  Computing (Allerton)}, Oct 2018, pp. 378--385.

\bibitem{kuang2020analysis}
Q.~Kuang, J.~Gong, X.~Chen, and X.~Ma, ``Analysis on computation-intensive
  status update in mobile edge computing,'' 2020.

\bibitem{DBLP:journals/corr/abs-1811-12924}
\BIBentryALTinterwordspacing
A.~O. Al{-}Abbasi and V.~Aggarwal, ``Joint information freshness and completion
  time optimization for vehicular networks,'' \emph{CoRR}, vol. abs/1811.12924,
  2018. [Online]. Available: \url{http://arxiv.org/abs/1811.12924}
\BIBentrySTDinterwordspacing

\bibitem{song2019age}
X.~Song, X.~Qin, Y.~Tao, B.~Liu, and P.~Zhang, ``Age based task scheduling and
  computation offloading in mobile-edge computing systems,'' \emph{arXiv
  preprint arXiv:1905.11570}, 2019.

\bibitem{rappaport1996wireless}
T.~S. Rappaport \emph{et~al.}, \emph{Wireless communications: principles and
  practice}.\hskip 1em plus 0.5em minus 0.4em\relax prentice hall PTR New
  Jersey, 1996, vol.~2.

\bibitem{chu2013heterogeneous}
X.~Chu, D.~Lopez-Perez, Y.~Yang, and F.~Gunnarsson, \emph{Heterogeneous
  Cellular Networks: Theory, Simulation and Deployment}.\hskip 1em plus 0.5em
  minus 0.4em\relax Cambridge University Press, 2013.

\bibitem{DBLP:journals/corr/abs-1807-04356}
\BIBentryALTinterwordspacing
B.~Zhou and W.~Saad, ``Joint status sampling and updating for minimizing age of
  information in the internet of things,'' \emph{CoRR}, vol. abs/1807.04356,
  2018. [Online]. Available: \url{http://arxiv.org/abs/1807.04356}
\BIBentrySTDinterwordspacing

\bibitem{zhang1999finite}
Q.~Zhang and S.~A. Kassam, ``Finite-state markov model for rayleigh fading
  channels,'' \emph{IEEE Transactions on communications}, vol.~47, no.~11, pp.
  1688--1692, 1999.

\bibitem{DBLP:journals/corr/abs-1710-04971}
\BIBentryALTinterwordspacing
E.~T. Ceran, D.~G{\"{u}}nd{\"{u}}z, and A.~Gy{\"{o}}rgy, ``Average age of
  information with hybrid {ARQ} under a resource constraint,'' \emph{CoRR},
  vol. abs/1710.04971, 2017. [Online]. Available:
  \url{http://arxiv.org/abs/1710.04971}
\BIBentrySTDinterwordspacing

\bibitem{altman1999constrained}
E.~Altman, \emph{Constrained Markov decision processes}.\hskip 1em plus 0.5em
  minus 0.4em\relax CRC Press, 1999, vol.~7.

\bibitem{puterman2014markov}
M.~L. Puterman, \emph{Markov Decision Processes.: Discrete Stochastic Dynamic
  Programming}.\hskip 1em plus 0.5em minus 0.4em\relax John Wiley \& Sons,
  2014.

\bibitem{bertsekas1995dynamic}
D.~P. Bertsekas, D.~P. Bertsekas, D.~P. Bertsekas, and D.~P. Bertsekas,
  \emph{Dynamic programming and optimal control}.\hskip 1em plus 0.5em minus
  0.4em\relax Athena scientific Belmont, MA, 1995, vol.~1, no.~2.

\bibitem{ross2014introduction}
S.~M. Ross, \emph{Introduction to stochastic dynamic programming}.\hskip 1em
  plus 0.5em minus 0.4em\relax Academic press, 2014.

\bibitem{robbins1951stochastic}
H.~Robbins and S.~Monro, ``A stochastic approximation method,'' \emph{The
  annals of mathematical statistics}, pp. 400--407, 1951.

\bibitem{6249269}
T.~{Soyata}, R.~{Muraleedharan}, C.~{Funai}, M.~{Kwon}, and W.~{Heinzelman},
  ``Cloud-vision: Real-time face recognition using a mobile-cloudlet-cloud
  acceleration architecture,'' in \emph{2012 IEEE Symposium on Computers and
  Communications (ISCC)}, July 2012, pp. 000\,059--000\,066.

\bibitem{tran2018joint}
T.~X. Tran and D.~Pompili, ``Joint task offloading and resource allocation for
  multi-server mobile-edge computing networks,'' \emph{IEEE Transactions on
  Vehicular Technology}, vol.~68, no.~1, pp. 856--868, 2018.

\end{thebibliography}
